\documentclass[prb,amsmath,amssymb]{revtex4}

\usepackage{graphicx}
\usepackage{dcolumn}
\usepackage{bm}
\usepackage{color}

\newcommand{\ve}[1][K]{\mathbf{#1}}

\def \equi#1{\mathrel{\mathop{\kern 0pt\sim}\limits_{#1}}} 

\begin{document}

\title{Everlasting impact of initial perturbations on first-passage times of non-Markovian random walks}

\author{N. Levernier$^1$,  T. V. Mendes$^{2}$, O. B\'enichou$^3$, R. Voituriez$^{3,4}$,  T. Gu\'erin$^{2}$}
\affiliation{$^{1}$ Aix Marseille Univ., Université de Toulon, CNRS, CPT, Turing Center for Living Systems, 13009 Marseille, France}
\affiliation{$^{2}$Laboratoire Ondes et Mati\`ere d'Aquitaine, University of Bordeaux, Unit\'e Mixte de Recherche 5798, CNRS, F-33400 Talence, France }
\affiliation{$^{3}$Laboratoire de Physique Th\'eorique de la Mati\`ere Condens\'ee, CNRS/UPMC, 
 4 Place Jussieu, 75005 Paris, France}
\affiliation{$^{4}$Laboratoire Jean Perrin, CNRS/UPMC, 
 4 Place Jussieu, 75005 Paris, France}

\bibliographystyle{naturemag.bst}

\begin{abstract}
\textbf{Persistence, defined as the probability  that a fluctuating signal has not reached a threshold up to a given observation time, plays a crucial role in the theory of random processes\cite{Majumdar1999,ReviewBray}. It quantifies the kinetics of processes as varied as phase ordering, reaction diffusion or interface relaxation dynamics. The fact that persistence can decay algebraically with time with non trivial exponents  has triggered  a number of experimental \cite{wong2001measurement,tam1997first,dougherty2002experimental,merikoski2003temporal,soriano2009universality} and theoretical \cite{Majumdar1996uq,Derrida1996,derrida1995exact,Majumdar1996,Majumdar2001fk,majumdar1996global,ehrhardt2002series,lee1997persistence,constantin2003infinite,Krug1997,Molchan1999} studies. However, general analytical methods to calculate   persistence exponents cannot be applied to the ubiquitous case of non-Markovian systems relaxing transiently after an imposed initial perturbation. Here, we introduce a theoretical framework that enables the non perturbative determination of  persistence exponents of $d$-dimensional Gaussian non-Markovian processes with general non stationary dynamics relaxing to a steady state after an initial perturbation.  Two  prototypical  classes of situations are analyzed:   either the system is subjected to a temperature quench at  initial time, or  its past trajectory is assumed to have been observed and thus known.  Altogether, our results reveal and quantify, on the basis of Gaussian processes, the deep impact of initial perturbations on first-passage statistics of non-Markovian processes. Our theory covers the case of spatial dimension higher than one, opening the way to characterize non-trivial reaction kinetics for complex systems with non-equilibrium initial conditions.} 
\end{abstract}

\maketitle

The persistence $S(t)$ is the probability that a random process $x(t)$ has not reached a threshold up to time $t$\cite{Majumdar1999,ReviewBray}. This quantity  is a natural tool  in non equilibrium statistical physics to probe  the history  of various systems undergoing phase ordering\cite{derrida1994non,derrida1995exact,Majumdar1996} or reaction diffusion dynamics\cite{ReviewBray}, or to quantify the efficiency of target search problems \cite{Redner:2001a,Condamin2007,Benichou2010,Schuss2007,godec2016universal,delorme2015maximum,guerin2016mean,levernier2019survival,dolgushev2015contact,Guerin2012a}. It has  been recognized that the long time decay of persistence  is often algebraic, $S(t)\sim t^{-\theta}$, where the  persistence exponent $\theta$ is non trivial as soon as the process is non-Markovian (i.e. displays memory effects). 
 
As a matter of fact, even for seemingly simple Gaussian dynamics where all correlation functions are known, $\theta$ is generally non-trivial and not known in closed form. This has triggered an intense theoretical activity for its determination. Existing approaches to quantify  persistence  exponents of Gaussian processes can be classified according to the nature, stationary or not, of the increments $x(t+\tau)-x(t)$. If these increments are stationary \textit{at all times}, meaning that their statistics do not depend on the  observation  
 time $t$ (such as in the case of the fractional Brownian motion), $\theta$ is exactly known \cite{Krug1997,Molchan1999,levernier2018universal}    $\ $\footnote{In $d$ dimensions, for scale-invariant processes with stationary increments \cite{levernier2018universal}, $\theta=1-Hd$}.
In the opposite case where the increments always depend on the   observation
time $t$  and thus never reach a stationary dynamics (i.e. are stationary \textit{at no times}), persistence exponents have  been calculated  for the specific cases of the random acceleration process \cite{BURKHARDT1993,deSmedt2001partial,poplavskyi2018exact} or  systems in which the dynamics occurs at zero temperature \cite{poplavskyi2018exact,dornic2018universal,derrida1994non,derrida1995exact,Majumdar1996,Majumdar1996uq,Derrida1996,watson1996persistence,newman2001critical}, and is thus deterministic with random initial conditions. 

However, numerous physical situations display a  relaxation dynamics  -- typically after an initial perturbation --  that becomes stationary only after a \textit{transient} regime. 
This is the rule for processes interacting with many degrees of freedom, subjected to thermal fluctuations during the dynamics, but prepared in a non-equilibrium   or perturbed state.   As a prototypical example, consider a tagged monomer of a flexible polymer initially equilibrated at a temperature  $T\not=1$, and  quenched to a different  temperature $T_0=1$ at time $t\ge0$. The dynamics of the tagged monomer  keeps transiently  track of this initial perturbation, and relaxes  to  the equilibrium state at $T_0$ with stationary increments. Persistence properties for such process with non stationary   increments (ie displaying aging), which in this example are instrumental to quantify the reaction kinetics of the polymer with a given reactive site, remain largely unknown. In fact, there is a fundamental reason why standard methods to calculate persistence exponents cannot be applied for transiently aging  processes (see SI, sections A and B)\footnote{More precisely, for transiently aging processes, the  independent interval approximation, which is usually applied to calculate the statistics of zero crossing of the Gaussian process obtained after Lamperti transform, cannot be applied since intervals between zero crossing become ill-defined, see SI Section A and B. }. The only available results for similar problems are limited to one-dimensional processes and  provide  bounds for the persistence exponents as well as perturbative expansions for weakly non-Markovian processes\cite{Krug1997}.

Here, we develop  a general theoretical framework that enables   the determination of  the persistence exponents of general  Gaussian processes displaying such  transient aging dynamics. We stress that these Gaussian processes are non-Markovian (display memory effects),  and appear in a wide range of contexts~\cite{Min2005,kou2004generalized,wei2000single,turiv2013,ochab2011scale,cutland1995stock,burnecki2012universal,ernst2012fractional,weiss2013single,mason1995optical}.
Our method  enables us to reveal and quantify the impact of initial conditions, such as a temperature quench,  on the persistence exponent. 
We also consider the case where the past trajectory of the random walker is known,    e.g. because it has been observed. We show that the very observation of this past trajectory modifies the persistence exponent which is quantified by our approach. 
Importantly, our theory covers the physically relevant and widely unexplored case of persistence for non-Markovian random walkers living in a space of dimension higher than one\footnote{Note that here the spatial dimension $d$ is different from the spatial dimension $D$ appearing in the problem of diffusive persistence\cite{Majumdar1996uq,Derrida1996}, where the process studied is $x(t)=\phi(\ve[0],t)$ where $\partial_t\phi=\nabla^2 \phi$, $\nabla$ being the $D$-dimensional nabla operator. Here, higher dimensions means that the variable $x$ it-self evolves in a $d$ dimensional space.}.

We first consider a one-dimensional isotropic non-Markovian Gaussian stochastic process $x(t)$, which represents the position of a random walker at time $t$. 
It is entirely defined by its mean value, assumed for simplicity to be constant with time (unbiased process), and its covariance ${\rm Cov}(x(t),x(t'))=\sigma_0(t,t')$. This covariance is assumed   to be given and to take the standard self-similar scaling\cite{ReviewBray} form at long times $t,t'\gg1$, $\sigma_0(t,t')\sim t^{2H}G(t/t')\equiv\sigma(t,t')$, where $H$ is the usual Hurst exponent. We chose our units of time so that $G(1)=1$. At long times, the mean square displacement $\sigma(t,t)= t^{2H}$  is assumed to diverge so that  the particle does not remain close to its initial position, which leads to $H>0$. 
Furthermore, we assume that the statistics of the increments $x(t+\tau)-x(t)$ become stationary at long times, i.e. become independent of the observation time $t$ when $t\to\infty$. 
This implies the existence of a transient regime associated to the progressive decay of the memory of the initial state, and defines a stationary covariance $\sigma_s$ given by
\begin{align}
\sigma_s(\tau,\tau')=&\lim_{t\to\infty} \langle [x(t+\tau)-x(t)][x(t+\tau')-x(t)]\rangle.  
\end{align}
Of note, the persistence exponent $\theta$ is  known to be given by $\theta=1-H$ under the stronger hypothesis that the statistics of the increments is stationary at \textit{any} time (\textit{i.e.} when $\sigma_s=\sigma_0$)  \cite{Krug1997,Molchan1999}. The class of random walks  that we consider here covers a broad spectrum of  non-Markovian processes used in physics, and in particular both subdiffusive ($H<1/2$) and superdiffusive ($H>1/2$) walks.

\textit{Theoretical method to determine $\theta$. } Our starting point is  the following  generalization of the renewal equation \cite{Redner:2001a}
\begin{align}
	p(0,t)=\int_0^t d\tau F(\tau) p(0,t\vert \mathrm{FPT}=\tau),\label{RenewalGeneralized}
\end{align}
which results from a partition over the first-passage event. In this equation, $p(0,t)$ stands for the  probability density that the random walker is at position $x=0$ at time $t$,  $F$ is the first-passage time (FPT) density  and  $p(0,t\vert \mathrm{FPT}=\tau)$ is the probability density that $x=0$ at time $t$ given that the first-passage event occurred at time $\tau$.

To proceed further, we   assume that the stochastic process in the future of the FPT, defined by $y(t)\equiv x(t+\mathrm{FPT})$, is  Gaussian with so far undetermined mean $\mu(t)$ and covariance  $\sigma_{\pi}(t,t')$. Such Gaussian approximation has proved successful to seize memory effects to predict mean first-passage times of Gaussian random walkers in confinement with stationary increments \cite{guerin2016mean,levernier2019survival,levernier2020kinetics};  in the present context simulations   show the broad validity of this hypothesis (see  SI, Fig.~S1). A first result of our approach is that the exponent $\theta$ is linked to the large time behaviour of $\sigma_{\pi}(t,t)$, which is found from Eq.~(\ref{RenewalGeneralized}) to behave like (see SI, Section D)
\begin{equation}
\sigma_{\pi}(t,t)-\sigma_0(t,t)\underset{t\to\infty}{\propto} \  t^{2H-\theta} \label{DefCorr}
\end{equation}
 This means that the calculation of the exponent $\theta$ amounts to that of the covariance $\sigma_{\pi}(t,t')$ of the trajectories in the late future of the first-passage.

Relying on a generalization of Eq.~(\ref{RenewalGeneralized})  to  link the two-time joint probability distribution functions of $x(t_1), x(t_2)$ and the FPT density, we obtain a self-consistent equation for the distribution of trajectories in the future of the FPT, leading in the large time limit to   (see SI, Section D for  details):
\begin{align}
\label{key}
 \int_0^\infty \frac{dt}{t^H}\Big\{&\rho(t+\tau, t+\tau')-\rho(t+\tau, t)\frac{\sigma(t+\tau',t)}{\sigma(t,t)}- 
 \rho(t+\tau', t)\frac{\sigma(t+\tau,t)}{\sigma(t,t)}\nonumber \\&
 +3\rho(t, t)\frac{\sigma(t+\tau,t)\sigma(t+\tau',t)}{2\sigma(t,t)^2}-\frac{\rho(t, t)}{2\sigma(t,t)}\left[\sigma(t+\tau,t+\tau') -\sigma_K(\tau,\tau') \right] 
\Big\}=0 
\end{align}
where $\rho\equiv \sigma_{\pi}-\sigma_0$ (for large times). Here, 
\begin{align}
\sigma_K(t,t')=\begin{cases}
\sigma(t,t')  & \text{if } \theta>1-H\\
\sigma_s(t,t')&\text{if } \theta<1-H
\end{cases}\label{DefSigmaK}
\end{align}
Next, we find that the linear equation \eqref{key} admits solutions of the scaling form $\rho(t,t')=t^{2H-\theta} z_\theta(t/t')$, where $z_\theta(u)$ satisfies a linear  integral equation of the form 
\begin{equation}
\label{eqint}
\int_0^1 K_\theta(u,v)\left[z_\theta(u)-z_\theta(1)\left(1-\frac{(2H-\theta) (1-u)}{2}\right)\right] {\rm d}u=f_\theta(v),
\end{equation}
where $K_\theta$ and $f_\theta$ are given in SI (Section D) in terms of $\sigma$.  It is found that generic solutions $z_\theta(u)$ display divergences for small $u$, and we argue that  $\theta$ is obtained by imposing that $z_\theta(u)$ is regular. We expect that this selection criterium is valid at least for $2H-\theta>0$ since it amounts in this case to impose that $\rho(t,0)=0$. Self-consistency reasons also lead us to restrict the analysis to $H>1/3$ (see SI, Section E).  In practice, the linear integral equation  Eq.~\eqref{eqint} is solved numerically for a test value  $\theta_{\rm test}$ and yields a diverging solution $z_{\theta_{\rm test}}(u)\sim A(\theta_{\rm test}) u^{-\alpha(\theta_{\rm test})}$; the persistence exponent $\theta$ is then obtained iteratively by enforcing that the prefactor vanishes, $A=0$ (see SI, Section D). This finally provides a constructive, non perturbative determination of the persistence exponent $\theta$ for  Gaussian process with general non stationary dynamics, which is the central result of this paper.

\begin{figure}
\includegraphics[width=14cm]{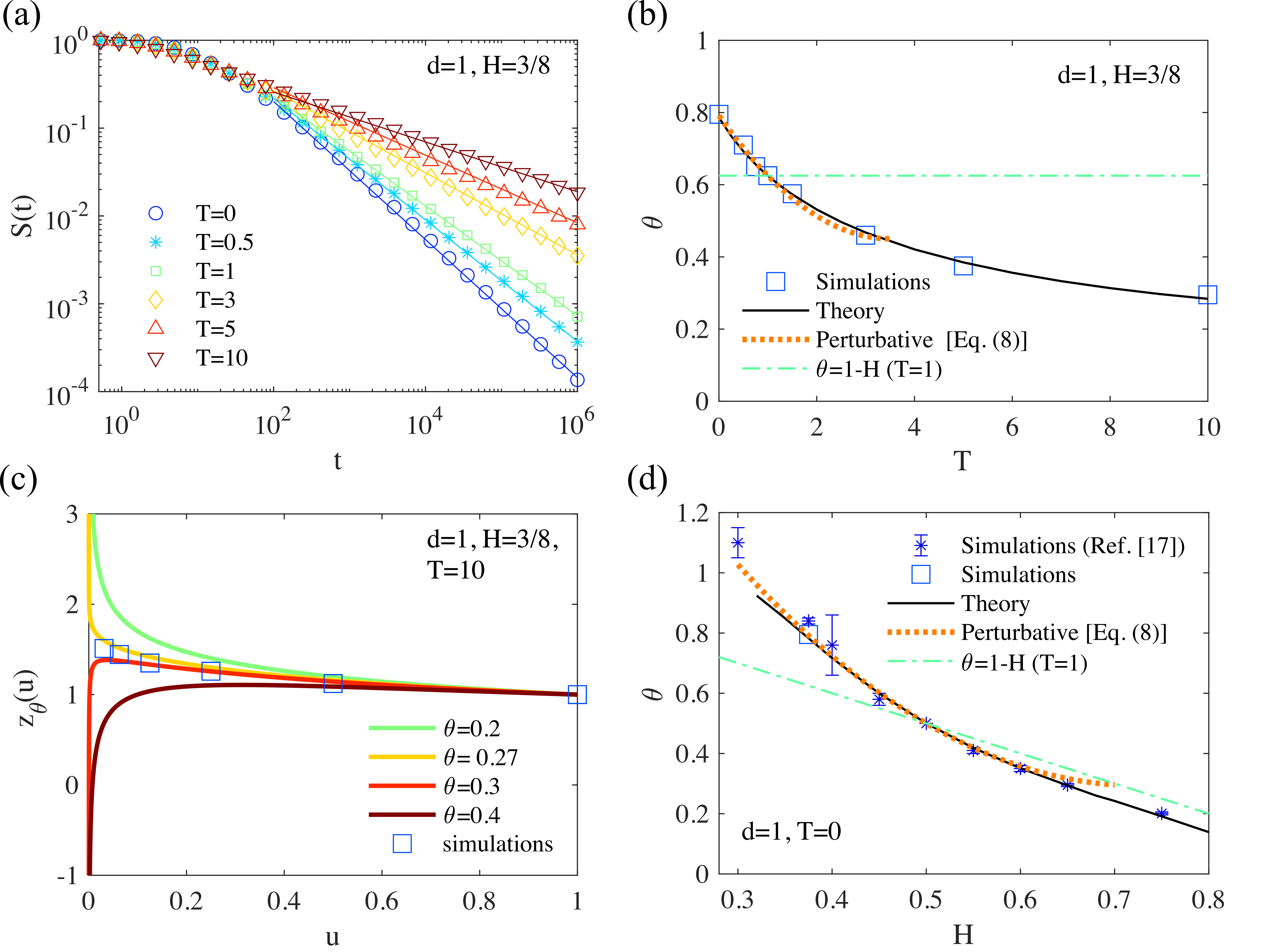} 
\caption{\textbf{Persistence for quenched fBM (type I problem) in $d=1$.} (a) Example of survival probabilities for interface dynamics with $H=3/8$ at different temperatures. The slopes of the continuous lines is the value of $\theta$ predicted in our approach. (b) Systematic comparison of $\theta$ as measured in simulations versus theoretical values for different $T$ and $H=3/8$. (c) Value of $z_\theta(u)$ as measured in simulations (squares) by analyzing the statistics of trajectories after the FPT, compared with theoretical values for different $\theta$. Notice the divergences for small $u$ towards $\pm\infty$, which enable us to select the value of $\theta$ to minimize these divergences. (d) Same as (b) for different $H$, with $T=0$ fixed. ``Star'' symbols are the simulation results of Ref. \cite{Krug1997}.}
\label{Fig1D}
\end{figure}

\textit{Applications. } We now show how these results enable us to determine the impact of initial conditions in two physically relevant cases. 
The first type of problems (called type I here) is the determination of $\theta$ in systems  which relax after a sharp temperature quench that occurs at initial time, which is a very generic situation. Typically, physical realizations of the random process $x(t)$ can be the position of a monomer in various models of macromolecules or the local height of an interface, which  span a number of values of $H$.  In all these models, assuming that
the initial state for $t\le0$ is an equilibrium state at temperature $T\neq1$, while the dynamics at $t>0$ occurs at  temperature   $T_0=1$, 
the covariance function for $t,t'>0$ takes the form (see Ref.\cite{Krug1997} and SI, Section C)
\begin{align}
\sigma(t,t')\propto T(t^{2H}+t'^{2H})+(1-T)(t+t')^{2H}-\vert t-t'\vert^{2H}.
\end{align} 
Of note, the temperature $T$ before the quench can be lower or larger than the temperature $T_0=1$ of the dynamics for $t>0$.
Examples of survival probabilities obtained from simulations are displayed in Fig \ref{Fig1D}(a), which clearly shows  that the persistence exponent depends on the choice of initial conditions, and that the dependence of the persistence exponent on  temperature is correctly predicted by our approach [Fig \ref{Fig1D}(b)]. Remarkably, the values of $\theta$ for different temperatures span a large set of values and are markedly different from their value $\theta =1-H$ in the stationary state. In one example of simulations, we also recorded the trajectories in the future of the first-passage and measured numerically the function $z_\theta(x)$, which shows good agreement with the theoretical prediction [see Fig \ref{Fig1D}(c)].  This figure also illustrates our procedure to determine   $\theta$ as defined above: the calculated $z_\theta(u)$ show divergence for small $u$   whenever $\theta$ is above or below its exact value. In Fig. \ref{Fig1D}(d), we check that our theory is also correct for different $H$ (focusing on $T=0$), for both superdiffusive and subdiffusive processes,  and even far from the Markovian regime $H=1/2$.  
In addition, explicit results can be obtained by analyzing our formalism [Eq.~(\ref{key})] perturbatively in the limit $\varepsilon=H-1/2\to0$. An expansion up to second order leads for any temperature $T$ before the quench  to 
\begin{align}
&\theta_{\mathrm{I}}= 1-H-2(\sqrt{2}-1)(1-T)(H-1/2) +\{a_1 \ (1-T)[a_2 +(1-T)] \}(H-1/2)^2+\mathcal{O}\left((H-1/2)^3\right),
\end{align}
where analytical expressions of $a_1$ and $a_2$ are given in SI (Section F), with numerical estimates $a_1\simeq1.77$, $a_2\simeq1.28$. Interestingly, in the particular cases $T=0$ and $T=1$, the first order terms coincide with the exact first order solution of  Ref.~\cite{Krug1997}, which points towards the exactness of our approach at this order.   These perturbative results are in good agreement with simulation results [Fig.~\ref{Fig1D}(c),(d)]. Finally, these results show that an imposed initial perturbation -- here a temperature quench, deeply impacts the  first-passage statistics of the system. In the case of subdiffusive (or antipersistent) dynamics ($H<1/2$, realized typically in polymer models), it is found that, because of long range memory effects,  an initial quench from a high ($T>1$) to a low ($T_0=1$) temperature can strongly slow down the first-passage kinetics ($\theta<1-H$), while a quench from low to high temperatures accelerates  the kinetics ($\theta>1-H$);  opposite conclusions are  reached for superdiffusive (or persistent) dynamics ($H>1/2$).

\begin{figure}[h!]
\includegraphics[width=14cm]{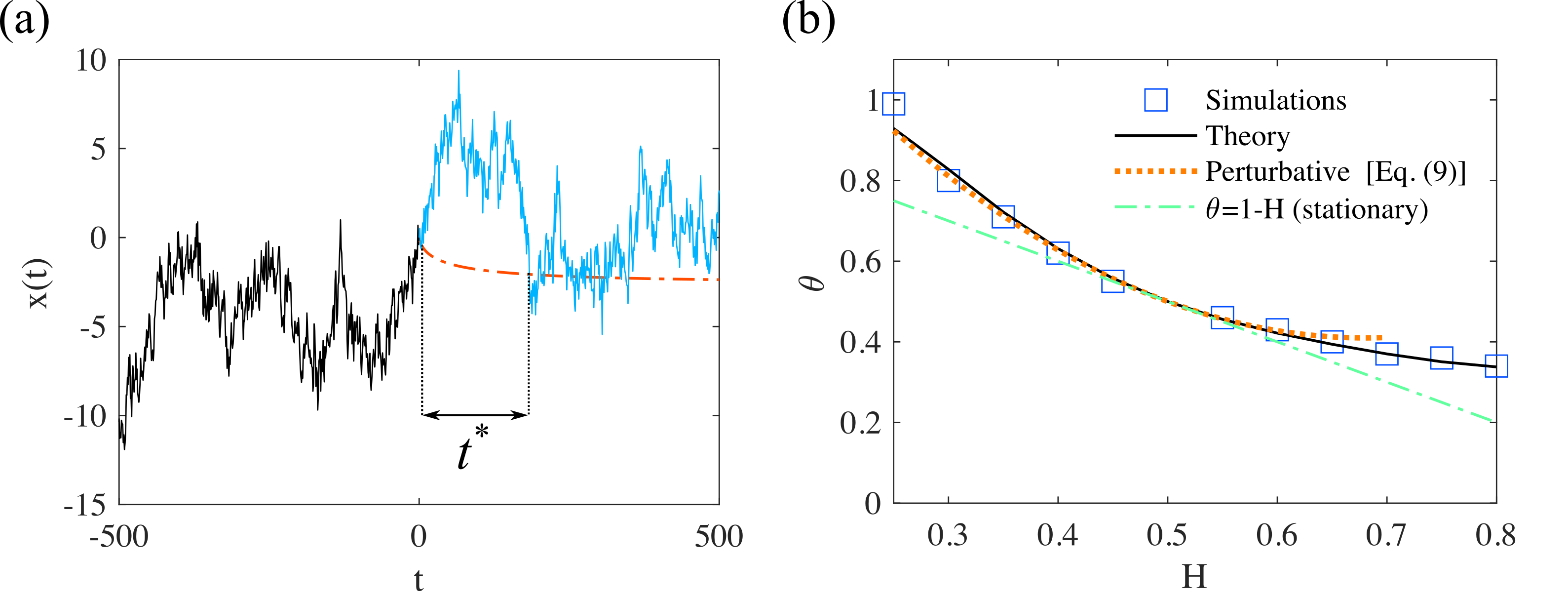} 
\caption{\textbf{Persistence for fBM conditioned on the past trajectory (type II problem)} (a) Definition of the problem: assume that a particular trajectory is observed for $t<0$ (black curve). The blue line is one realization of trajectory for $t>0$, the dashed red line represents the  average trajectory given that the past trajectory is observed. The time $t^*$ is the first crossing time  to this predicted average trajectory, and the persistence exponent characterizes the probability that $t^*>t$, for large $t$. (b) Comparison between values of $\theta$ for the fBM conditioned on the past trajectory, obtained in simulations (symbols), our theoretical approach (black line) and perturbation expansion (dashed red line). We also indicate the value   $\theta=1-H$ for non-conditioned fBM.  }
\label{Fig1DPast}
\end{figure}

The second class of problems (type II) corresponds to the determination of $\theta$   in an idealized situation where a given trajectory for $x(t<0)$ is assumed to be accessible  and  observed at all  times $t<0$ ; this can be realized in various settings, ranging from single particle tracking techniques in the context of transport in complex systems, to the  monitoring of the value of an asset in the context of financial markets. Here we aim at quantifying the impact of such observation of the system in the past ($t<0$) on its future dynamics ($t>0$).   
It is known  \cite{yaglom1955correlation,GRIPENBERG1996,anh2004prediction,inoue2012prediction} that the mean future trajectory (for $t>0$), conditional to a given observation in the past $x(t<0)$, can be expressed as a linear combination of all   positions in the past.  
Our approach makes it possible to  determine quantitatively the exponent $\theta$ characterizing  the probability $S(t)$ of not crossing this  average conditional trajectory, or of not reaching a fixed threshold above (or below) it, see Fig.~\ref{Fig1DPast}(a). Strikingly, we find that the value of $\theta$ can be significantly larger than the value $\theta=1-H$ obtained in absence of any prior observation of the system. It  does not depend on the particular realization of the observed past trajectory, but only  on the fact that this observation is available. This thus shows that the very observation of the system can drastically impact the future first-passage statistics, and in fact effectively accelerate the dynamics at large times because  $\theta\ge 1-H$ for all values of $H$, irrespective of the persistent or antipersistent nature of the process. The results in Fig.~\ref{Fig1DPast}(b) show again a good agreement between the predicted values of $\theta$ and simulations. As above a perturbation expansion of our formalism can be performed for weakly non-Markovian processes, leading to the explicit result
\begin{align}
\theta_{\mathrm{II}}=1-H+4 \ln 2 (H-1/2)^2+ \mathcal{O}((H-1/2)^3),
\end{align}
which is supported by our simulations [Fig.~\ref{Fig1DPast}(b)]. 

\begin{figure}[h!]
\includegraphics[width=14cm]{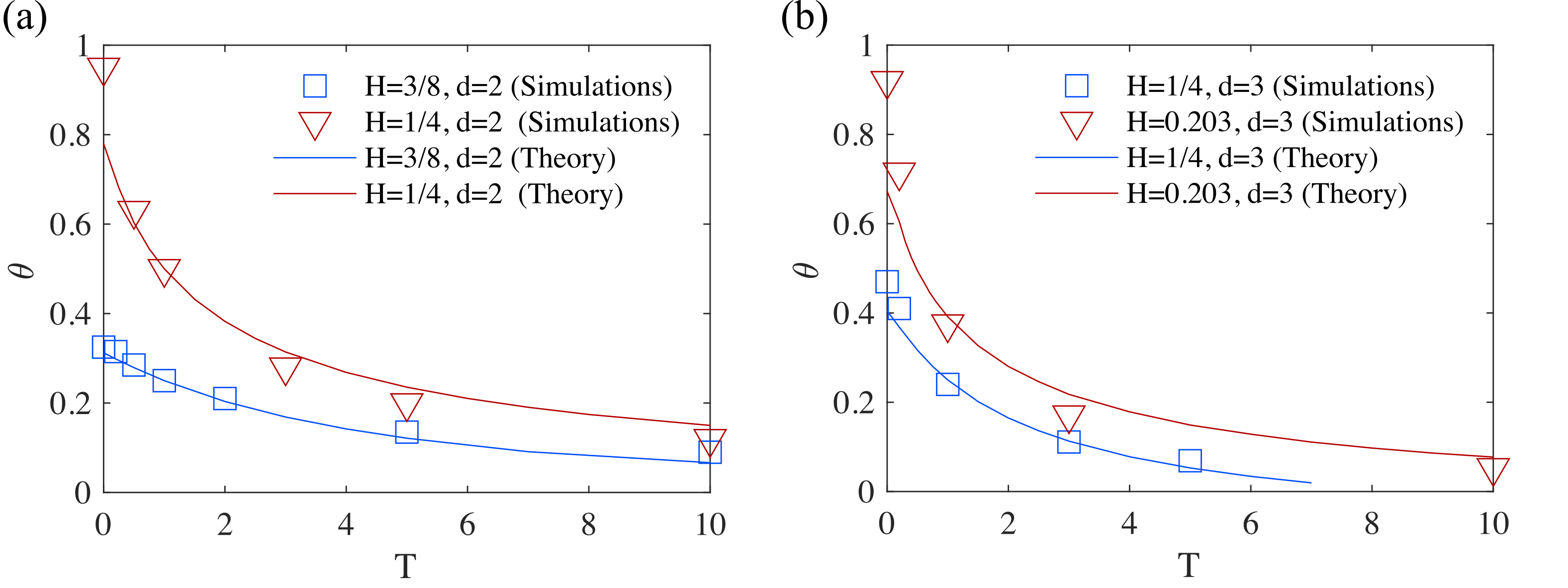} 
\caption{\textbf{Persistence exponents in dimensions higher than one} (a) $d=2$ and (b) $d=3$. These results are obtained when the random walker is a tagged monomer in  different polymer models: semi-flexible chains ($H=3/8$), flexible chain without hydrodynamic interactions ($H=1/4$), macromolecule of fractal architecture (Vicsek fractal, $H\simeq 0.203$).}
\label{Fig2D3D}
\end{figure}

\textit{Persistence in higher dimensions. }
Our theory can be generalized to the case of an isotropic Gaussian random process $\ve[x](t)$ evolving in a space of dimension $d>1$. In this case, to define the survival probability we replace the condition of reaching a threshold by  the condition of reaching a target.  To the best of our knowledge, in this case the persistence exponent has not been investigated in the  literature for non-Markovian walks with non-stationary initial conditions, despite its obvious relevance to reactivity problems in complex systems. Here we restrict ourselves to the case where a target, even point-like, is found with probability one (compact case, when $dH<1$).  It turns out that very few changes are needed to generalize the theory in $d$ dimensions, generalized versions of the equations are presented in SI (Section C), and we restrict ourselves to $H>1/(2+d)$. 
Fig. \ref{Fig2D3D} shows simulation results when $\ve[x](t)$ is the position of a monomer in various polymer models: semi-flexible or flexible  chains, or fractal hyperbranched flexible macromolecules.  It is found that our theory captures quantitatively the dependence  of the persistence exponents on the temperature quench for all these models.   This dependence on the temperature quench shows that the exponents describing the kinetics of absorption to a target is  significantly modified by preparing the system with non-stationary initial conditions. For example, if the random walker is a tagged monomer of a macromolecule, a non-equilibrium condition could be obtained by   a temperature quench, or by imposing a constraint, such as a geometric confinement or an external field,  that is relaxed at $t=0$. Alternatively, if memory effects of the random walker come from its interactions with a surrounding viscoelastic medium, a non-equilibrium initial state could be obtained by imposing a sharp change of the parameters characterizing this medium.   In these cases, we predict that the reaction kinetics to a target can be deeply impacted, and  display non-trivial exponents quantified by our approach (as soon as the covariance $\sigma$ can be calculated).

\textit{Conclusion.} In Fig.~\ref{FigAllTheta}, simulation data for $\theta$ for all the models considered in this work are recapitulated and  compared with the values predicted by our approach. The data collapse shows an excellent agreement and validates our method.  The slight departures of simulations from theory  occur only when $2H-\theta>0$ (see Fig.~\ref{FigAllTheta}), in agreement with   our previous remark that our selection criterium may not be valid anymore in this regime. Similarly, the curves which are the least precise on Fig.~\ref{Fig2D3D} are those for which $H$ is very close to $1/(2+d)$, where the theory is not expected to give accurate results anymore.  Altogether, this shows that our theory  provides a non perturbative, constructive, quantitative determination of the persistence exponents for general   Gaussian stochastic processes with non-stationary initial conditions, which typically model the relaxation after an initial perturbation of  systems with non Makovian dynamics, such as tracer particles in complex environments with many interacting  degrees of freedom. Our results demonstrate  that  initial perturbations can have a deep, long lived  impact on the  first-passage statistics of non-Markovian processes. Importantly, our theory also predicts non-trivial exponents in dimension higher than one, and thus opens the way to the quantification and control of  reaction kinetics for complex systems with non-equilibrium initial conditions.

\begin{figure}[h!]
\includegraphics[width=7cm]{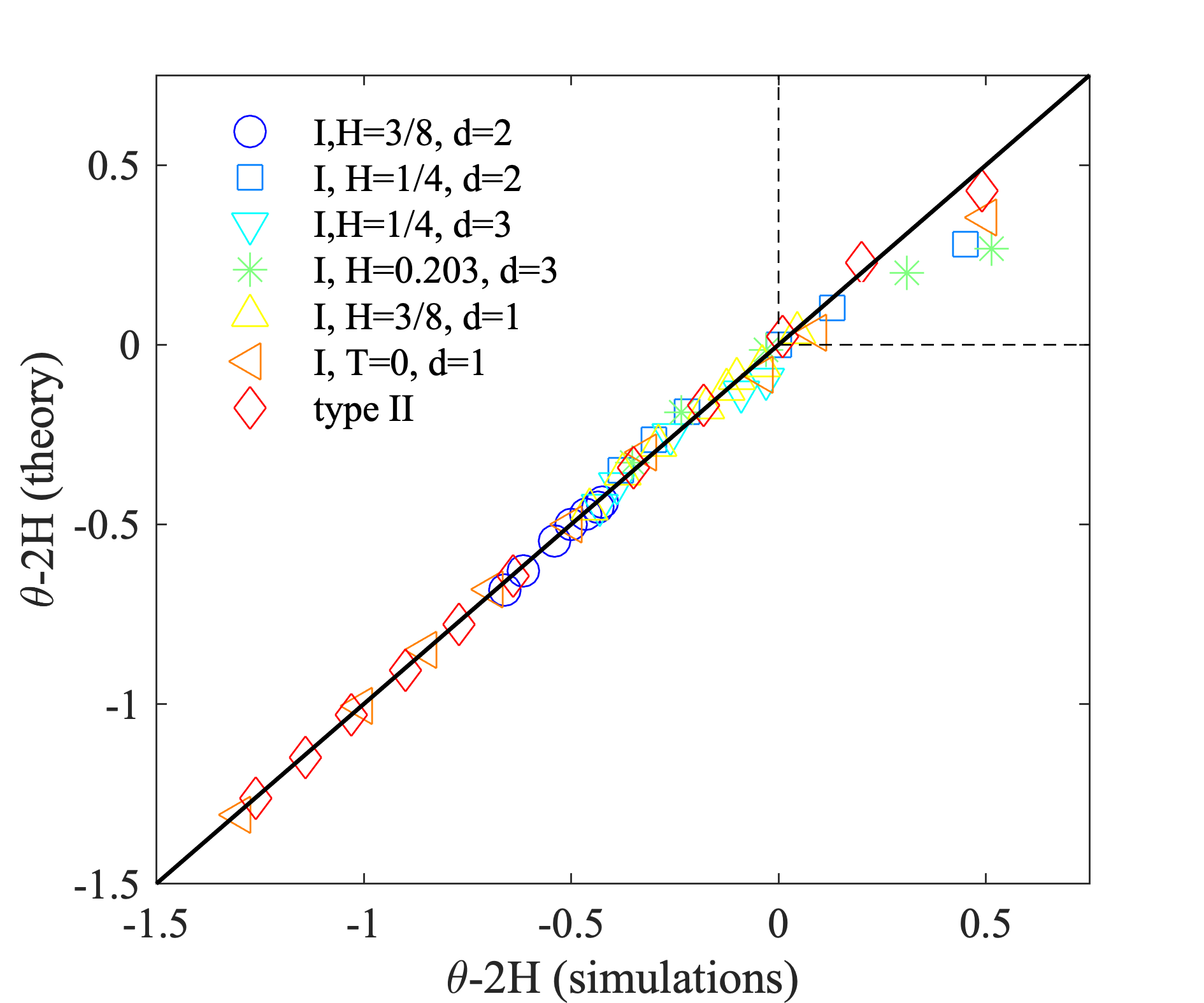} 
\caption{\textbf{Comparison of theoretical values of the persistence exponent  versus simulations} for all data shown in Figs. \ref{Fig1D}, \ref{Fig1DPast}, \ref{Fig2D3D}. }
\label{FigAllTheta}
\end{figure}

 \begin{acknowledgments}
Computer time for this study was provided by the computing facilities MCIA (Mesocentre de Calcul Intensif Aquitain) of the Universit\'e de Bordeaux and of the Universit\'e de Pau et des Pays de l’Adour.  T-V. M. and T. G. acknowledge the support of the grant \textit{ComplexEncounters}, ANR-21-CE30-0020-01.
\end{acknowledgments}

\newpage 

\appendix 
\begin{center}
\Large\textbf{Supplementary Information}
\end{center}

In this Supplementary information, we provide:
\begin{itemize}
\item a characterization of Gaussian processes that have stationary increments at long times (Section \ref{SectionIncrLongTimes}),
\item the reason why persistence exponents for such processes cannot be obtained by standard methods (Section \ref{SectionFailIIA}),
\item the list  of the stochastic processes (and associated simulation algorithms) investigated in this work  (Section \ref{SectionListProcesses}), 
\item a detailed description of the derivation of Eqs. (3), (4), (6) of the main text (Section \ref{FormalismDetail}), including the description of how  $\theta$ is numerically obtained (Section \ref{SelectTheta}).
\item a discussion on the restrictions of the theory (Section \ref{SectionValidity}),
\item a perturbative analysis of our theory around the Markovian Brownian motion ($H\to1/2$, Section \ref{PertSection}),
\item the analysis of the quenched fBm in the limit of infinite initial temperature (Section \ref{InfiniteT}). 
\end{itemize}

\section{Characterization of Gaussian processes with stationary increments at long times}
\label{SectionIncrLongTimes}
We consider a one-dimensional Gaussian random walk  $x(t)$ evolving with continuous time $t\ge0$. Since the process is Gaussian, it is entirely defined by its mean $\langle x \rangle $, which we assume to be constant with time (symmetric walk), and by its covariance $\mathrm{Cov}(x(t),x(t'))=\sigma_0(t,t')$. At long times, we assume that $\sigma_0(t,t')$ admits the  scaling behavior:
\begin{align}
\sigma_0(t,t')\underset{t,t'\to\infty}\sim t^{2H}\ G(t/t')\equiv \sigma(t,t') , \label{ScalingSigma0}
\end{align}
where $G$ is a scaling function, $H$ is the Hurst exponent, and the symbol $\sim$ represents the mathematical asymptotic equivalence. The above scaling behavior is standard (see e.g. Ref.~\cite{ReviewBray}). 
Note also that $\sigma(t,t')$ is the covariance function at long time scales and is defined by Eq.~(\ref{ScalingSigma0}). We will restrict our-selves to $H>0$ (so that the particle is not trapped at long times at any position of space), and also to $H<1$ (the long-time process is then non-smooth \cite{ReviewBray}).  

In this work, we study Gaussian stochastic processes that have stationary increments \textit{at long times}, thus displaying \textit{transient} aging. 
In other words, we assume that the statistics of increments $x(t+\tau)-x(t)$ do not depend on the time elapsed  $t$, if $t$ is large enough. For such processes, one can define a ``stationary covariance'' $\sigma_s$ by: 
\begin{align}\label{DefSigmaS}
\sigma_s(\tau,\tau')=&\lim_{t\to\infty} \langle [x(t+\tau)-x(t)][x(t+\tau')-x(t)]\rangle.
\end{align}
For Gaussian processes whose covariance satisfies Eq.~(\ref{ScalingSigma0}), the existence of   $\sigma_s$ requires conditions on the behavior of the scaling function $G$ which we determine now. Using Eq.~(\ref{ScalingSigma0}), we obtain
\begin{align}\label{DefSigmaS}
\sigma_s(\tau,\tau')=&\lim_{t\to\infty} \left\{(t+\tau)^{2H} G\left(\frac{t+\tau'}{t+\tau}\right)-t^{2H}\left[G\left(\frac{t}{t+\tau}\right)+G\left(\frac{t}{t+\tau'}\right)-G\left(1\right)\right] \right\}
 \end{align}
Thus, the quantity $\sigma_s$ exists when the behavior of $G(u)$ near $u=1$ is  
\begin{align} 
G(u)\underset{u\to1, u<1}{=} G(1)[1+H(1-u)]-\kappa_s (1-u)^{2H} +  \text{h.o.t}  \label{BehaviorGSummary},
\end{align}
which holds for $u<1$. Here, $\text{h.o.t}$ means higher order terms, and $\kappa_s$ is a positive constant. For $u>1$ the behavior of $G$ is deduced from the condition that $\sigma(t,t')=\sigma(t',t)$ is symmetric. Note that specifying the linear term (proportional to $1-u$) is not necessary for $H<1/2$ but it dominates for superdiffusive processes so that it is important in this case. In fact, all correlators we will study in this work satisfy the above condition, irrespectively on the value of $H$. 
With the above condition,  the stationary covariance reads
\begin{align}
\sigma_s(\tau,\tau') =\kappa_s\left(\ \vert\tau\vert^{2H}+\vert\tau'\vert^{2H} - \vert\tau'-\tau\vert^{2H}  \right)= \tau^{2H} G_s(\tau/\tau'). \label{DefSigmaS}
\end{align}
This means that the increments in the future of $t$ have the same covariance function as that of the fractional Brownian motion (fBm), multiplied by a generalized transport coefficient $\kappa_s$. Here, $G_s$ is the scaling function associated to the stationary state and is defined by the above equation. The above formula holds also when $\tau,\tau'$ are negative. 

Alternatively, to define the    stationary property at long times, we could  require that, if we look only at trajectories satisfying the constraint $x(t)=x^*$, and that one looks at the trajectories in the future of this time $t$, then the statistics of these conditioned trajectories do not depend on $t$, if $t$ is large enough. In other words, we could define  $\sigma_s(\tau,\tau')$  as the limiting value of the covariance of $x(t+\tau)$ and $x(t+\tau')$ \textit{given that $x(t)=x^*$ is fixed}, when $t$ is large enough: 
\begin{align}
\sigma_s(\tau,\tau')=\lim_{t\to\infty}\mathrm{Cov}(x(t+\tau),x(t+\tau')\vert x(t)=x^*)=\lim_{t\to\infty} \left\{\sigma_0(t+\tau,t+\tau')-\frac{\sigma_0(t+\tau,t)\sigma_0(t,t+\tau')}{\sigma_0(t,t)}\right\} \label{DefSigmaStar},
\end{align}
where the notation $\text{Cov}(A,B\vert E)$ represents the covariance of $A,B$ given that the event $E$ is realized, and the second equality is deduced from general formulas on conditional Gaussian processes (see e.g. Ref. \cite{Eaton1983}). 
It turns out that the  condition for the existence of $\sigma_s$ defined in    (\ref{DefSigmaS}) or (\ref{DefSigmaStar}) is the same and is given by Eq.~(\ref{BehaviorGSummary}), and that the two definitions of $\sigma_s$ are identical. Hence, requiring that the increments of the trajectories are stationary at long time, or  imposing a condition on the stationarity of conditional covariances, is actually the same. 
It is instructive to consider the   convergence of the conditional covariance towards the stationary covariance:
\begin{align}
\mathrm{Cov}(x(t+t_1),x(t+t_2)\vert x(t)=0)-\sigma_s(t_1,t_2)\underset{t\to\infty}{\sim} - \frac{\kappa_s^2 (t_1t_2)^{2H}}{t^{2H} G(1)}\label{ConvergenceToStatCov}.
\end{align}
This convergence is therefore algebraic, which was expected due to the absence of characteristic relaxation time.

In summary, in this Section we have characterized the Gaussian processes that have long time stationary increments: these are the Gaussian processes that display a scaling function $G$ satisfying the condition (\ref{BehaviorGSummary}).

\section{The reason why standard methods to calculate persistence exponents fail when applied to Gaussian processes with stationary increments at long times}
\label{SectionFailIIA}

For one-dimensional Gaussian stochastic processes, a standard method to calculate the persistence exponent $\theta$ (defined so that $S(t)\propto 1/t^{\theta}$ at long times) consists in applying the Lamperti transform to obtain a stationary Gaussian process, and then to apply the independent interval approximation (IIA). This method is not exact but gives in general very good estimates of   persistence exponents \cite{Krug1997}. The Lamperti transform consists in defining logarithmic times $T \equiv \ln t$ and rescaling $x(t)$ by its root mean square to obtain a stationary process, i.e. defining the stochastic process $X(T)\equiv x(t)/[G(1)t^{2H}]^{1/2}$. The covariance function of the transformed process $X(T)$ (still Gaussian) reads  
\begin{align}
\langle X(T_1)X(T_2)\rangle=\frac{t_1^{2H}G(t_1/t_2)}{G(1)t_1^{H}t_2^{H}}=e^{-H(T_2-T_1)}\frac{G(e^{-(T_2-T_1)})}{G(1)}\equiv a(\vert T_2-T_1\vert).
\end{align}
This correlation function depends only on the difference $T_2-T_1$, meaning that the  process $X(T)$ is stationary. For small positive $T$, with the hypothesis (\ref{BehaviorGSummary}), this correlation function behaves as
\begin{align}
a(T)\underset{T\to0}\simeq (1-HT+...)[G(1)(1+HT)-\kappa_s T^{2H}+...]/G(1)=1-[\kappa_s/G(1)] T^{2H} + \text{h.o.t}.\label{Behavior_a}
\end{align}
For $0<H<1$, the above expansion means that $X(T)$ is a \textit{non-smooth} Gaussian process, because  smooth Gaussian processes admit a correlation function behaving as  $a(T)=a(0)-b T^2+...$ for $T\to0$. Now, the IIA method consists in evaluating the statistics of the time between two successive zeros of a smooth stationary process, but by construction for a non-smooth process this very concept of interval between two zeros is ill-defined. Hence by construction the IIA method can be applied only to smooth Gaussian processes. Therefore it cannot be applied to evaluate persistence exponents in our problems, where the dynamics displays stationary increments at long times. This has been noted in Ref.  \cite{Krug1997} for the study of persistence in fluctuating interfaces, but what we have shown here is that this restriction is more general and associated to the fact that the process has stationary increments at long times.

\section{List of stochastic processes   investigated in this work}

\label{SectionListProcesses}

\subsection{``Quenched'' fBm (type I process)}
We describe here the stochastic processes for which we will study persistence exponents. We will consider  a process which we call    ``quenched'' fBm, for which the correlator at long times is 
\begin{align}
\sigma(t,t')=\sigma_\text{I}(t,t')\equiv\frac{(1-T )(t+t')^{2H}+ T  [t^{2H}+t'^{2H}]-\vert t-t'\vert^{2H} }{[(1-T )2^{2H}+2T ] }.\label{sigmaQuenched}
\end{align}
The function $G$ associated to this correlator satisfies the condition (\ref{BehaviorGSummary}), meaning that it corresponds to a process $x(t)$ that becomes with stationary increments at long times. 
Depending on $H$, this correlator corresponds to various physical models (which we used to simulate $x(t)$ for various values of $H$). For example, in $d=1$, $x(t)$ can be seen as the local height of a one dimensional interface $x(t)=h(s=0,t)$ which obeys the dynamics
\begin{align}
\partial_t h=-(-\partial_s^2 )^{1/(2-4H)} h(s,t)+\xi(s,t), \hspace{1cm} \langle \xi(s,t)\xi(s',t')\rangle=2T_1 \delta(s-s') \delta(t-t') \label{DynInterface},
\end{align}
where $T_1$ is the temperature during the dynamics. Here, we will focus on the cases $H=1/4$ (Edwards-Wilkinson dynamics) and $H=3/8$ (Mullins-Herring dynamics). The correlator (\ref{sigmaQuenched}) is obtained when $x(t)=h(s=0,t)$ is calculated with an initial condition corresponding to a stationary state of Eq.~(\ref{DynInterface}) with initial temperature $T_1=T$, while the dynamics takes place at temperature $T_1=1$. The fact that (\ref{sigmaQuenched}) is obtained after a sharp change of temperature in a model of collective dynamics justifies the name ``quenched fBm''. Interestingly, the  correlator (\ref{sigmaQuenched}) with $T=0$ and $H=1/4$ is also that of a tagged particle in single file diffusion in crowded narrow channels, for which the position of all other particles is initially fixed (which corresponds to the zero temperature state in this problem) \cite{krapivsky2015dynamical}. 
Next, if one replaces $h(s,t)$ by $\ve[r](s,t)$ in Eq.~(\ref{DynInterface}), one obtains the equation for the dynamics of the positions of monomers $\ve[r](s,t)$ at time $t$ and position $s$ along the chain, for the Rouse chain ($H=1/4$, a bead-spring chain) or the semi-flexible chain ($H=3/8$).
Other values of $H$  can be obtained for the dynamics of the position $\ve[r]_i(t)$ of monomers in hyperbranched (fractal) macromolecules 
\begin{align}
\partial_t \ve[r]_i=-\sum_{j}A_{ij} \ve[r]_j(t) + \ve[f](s,t), \hspace{1cm} \langle f_{i,\alpha}(s,t)f_{j,\beta}(s',t')\rangle=2 T_1 \delta_{ij} \delta(t-t')\delta_{\alpha,\beta} \label{DynVF},
\end{align}
where $\alpha,\beta$ are spatial coordinates ($x,y,z$ in $d=3$) and $A_{ij}$ is the dynamical matrix, with $A_{ij}=-1$ if beads $i,j$ are connected and zero in the contrary, and $A_{ii}$ is the number of connected neighbors around bead $i$.  

\textbf{Simulations.- } In practice, we have simulated processes $x(t)$ (or $\ve[r](t)$ in $d$ dimensions) by integrating numerically a discretized versions of Eq.~(\ref{DynInterface}) (for $H=1/4$ and $H=3/8$) or Eq.~(\ref{DynVF}), for which we took the connectivity matrix of a Vicsek fractal of functionality $f=4$ \cite{blumen2004generalized}, with $H=\ln(3)/\{2 \ln[3(1+f)]\} \simeq 0.203$. Initially the polymer chain (or interface) was prepared at temperature $T$ with a reactive monomer at distance $x_0$ from the target and we recorded the time to the crossing of $x=0$ (in $d=1$), or the time to reach a sphere of radius $a$ in $d$ dimensions. To simulate the dynamic of VFs, we have decomposed the dynamics on eigenmodes
\begin{align}
\ve[r]_n=\sum_i b_i \ve[a]_i \ ; \  \partial_t \ve[a]_i=-\lambda_i \ve[a]_i+\ve[f]_i(t), \langle f_i(t)f_j(t')\rangle=2T_1\delta(t-t')\delta_{ij}\label{EqEigenmode}
\end{align}
where $\ve[r]_n$ is the reactive monomer, $\lambda_i$ are the eigenvalues of $A$ and $b_i$ the projection of the vector $(0,...,0,1,0,...0)$ on the eigenspace associated to $\lambda_i$. We used the   methods of Ref \cite{dolgushev2015contact} to identify iteratively the set of $(\lambda_i,b_i)$. We simulated  (\ref{EqEigenmode}), which is much more efficient than using (\ref{DynVF}) since the number of distinct eigenvalues for fractals is much smaller than the number of beads. In all simulations, we measured $\theta$ by plotting the survival probability in log-log scale, including the largest times allowed by the number of accumulated stochastic trajectories. Note that, in particular for low values of $\theta$, the regime of low values of $S(t)$ could not be reached so that the measurement of $\theta$ is not very accurate. We checked that the measured values do not depend on microscopic parameters (initial distance to target, target size, time steps, number of monomers...).

\subsection{fBm constrained on its past (type II process)}

We consider here the statistics of trajectories of a fractional Brownian motion conditioned on its past value. This process is obtained by considering a standard fractional Brownian motion, and asking for the statistics of $x(t)$ for $t>0$, given that the past trajectory $x(\tau)$ is known exactly for all past times $\tau<0$. To understand how this process is built, let us consider the case that the process is conditioned on $n$ past times $\tau_i<0$ (instead of all past times). In this case, general formulas on Gaussian processes (see e.g. chapter 3 in Ref \cite{Eaton1983}) show that the statistics of the conditioned process can be obtained from the knowledge of the covariance matrix $S_{ij}=\sigma_s(\tau_i,\tau_j)$  (for the unconditioned process, at the times $\tau_i$ at which one imposes the conditions):
\begin{align}
&\mathbb{E}(x(t) \vert [x(\tau_1),...x(\tau_n)] )=\sum_{i,j=1}^n\sigma_s(t,\tau_i) \ (S^{-1})_{ij} \ x(\tau_j), \label{643}\\
&\mathrm{Cov}(x(t),x(t')\vert  [x(\tau_1),...x(\tau_n)])=\sigma_s(t,t')-\sum_{i,j=1}^n\sigma_s(t,\tau_i) \ (S^{-1})_{ij}\ \sigma_s(t',\tau_j). \label{644}
\end{align}
The conditional means and covariances can thus be found at the cost of solving a linear equation to find the inverse matrix $S^{-1}$. In the continuous  limit $n\to\infty$, the problem of finding $S^{-1}$ is replaced by an integral equation  which was solved by Yaglom (for $H<1/2$) \cite{yaglom1955correlation} and Gripenberg and Norros (for $H>1/2$) \cite{GRIPENBERG1996} (see also \cite{anh2004prediction,inoue2012prediction}), with the result: 
\begin{align}
\mathbb{E}(x(t) \vert [x(\tau), \tau<0] )=\begin{cases}
 \frac{\cos(\pi H)}{\pi}\int_{0}^\infty d\tau \left(\frac{t}{\tau}\right)^{H+1/2} \frac{x(-\tau)}{t+\tau} & (H<1/2),\\
x(0)+ \frac{\cos\pi H}{\pi}\int  dx(-\tau) \int_0^{t/\tau} dv \frac{v^{H-1/2}}{v+1} & (H>1/2).
\end{cases}\label{RO431}
\end{align}
Note that in the second line the integration runs over stochastic trajectories, see Ref.~\cite{GRIPENBERG1996}. 
It is clear from  Eqs. (\ref{643}), (\ref{644}) that the formula for the conditional covariance is the same as that for the conditional mean (up to an additive term $\sigma_s$) if we replace the past trajectory $x(\tau_i)$ by $\sigma_s(t',\tau_i)$, so that   
\begin{align}
\sigma(t,t')=\text{Cov}(x(t),x(t') \vert [x(\tau), \tau<0] )= \sigma_s(\tau,\tau')-\frac{\cos(\pi H)}{\pi} \times \begin{cases}
\int_{0}^\infty d\tau \left(\frac{t}{\tau}\right)^{H+1/2} \frac{\sigma_s(t',-\tau)}{t+\tau} & (H<\frac{1}{2})\\
 \int_0^\infty d\tau  \int_0^{t/\tau} dv \frac{v^{H-1/2}}{v+1}\frac{d}{d\tau}\sigma_s(t',-\tau) & (H>\frac{1}{2})
\end{cases} 
\end{align}
Note that, in the case $H>1/2$ in Eq.~(\ref{RO431}) we have considered that the integration over the trajectory can be replaced by a sum over $x(t_{i+1})-x(t_i)$ in discrete time, applying the same linear operator on $\sigma_s$ leads to the derivative of $\sigma_s$ with respect to $\tau$ in the above equation. Replacing $\sigma_s$ by its value (\ref{DefSigmaS}), we obtain the following formula for the covariance of a fBm conditioned on its past:
\begin{align}
\sigma(t,t')=\sigma_\text{II}(t,t')\equiv  \kappa_s \left\{-\vert t-t'\vert +t^{2H}+t'^{2H}+\int_0^\infty dx \frac{\cos(\pi H)\ [(t+x t')^{2H}-t^{2H}-(t'x)^{2H}]}{\pi \ x^{H+1/2}(1+x)} \right\}\label{EqSigmaPast},
\end{align}
which is valid both for $H>1/2$ and $H<1/2$. 
Note that the normalization $\sigma(t,t)=t^{2H}$ is obtained if we chose
$\kappa_s= \Gamma(1-H)\ \Gamma(1/2+H)/[4^H \sqrt{\pi}]$

\textbf{Simulations.} To simulate a fBm constrained on its past, we used the Hosking algorithm \cite{hosking1984modeling,Dieker2004} in which the fBm is iteratively sampled at fixed time steps $t_i=n\Delta t$, each position $x(t_{i+1})$ being generated from $x(t_0=0),x(t_1),...,x(t_i)$. To take into account the fact that the past trajectory is known, we defined a time $t_m$ and replaced all positions $x(t_i)$ with $t_i<t_m$ by a starting position $x_0$. The next steps $x(t_j)$ with $t_j>t_m$ were generated according to the standard Hosking algorithm.  In the limit $t_m\to\infty$, this corresponds to a fBm constrained on its past trajectory after removing its average value to obtain a symmetric (centered) process.

\section{Detailed derivation of the equations of the formalism}
\label{FormalismDetail}
 
\subsection{The link between the covariance of trajectories in the late future of the first-passage time and the persistence exponent [Derivation of Eq.~(3)]}

Here, we consider a stochastic process in $d$ dimensions, $\ve[x](t)=(x(t),y(t),...)$. We suppose it is Gaussian, so that it is defined by its mean value and its covariance. We also assume that it is isotropic, so that the mean value $\langle \ve[x](t)\rangle$ is constant with time, and the covariance   between coordinates takes the form $\text{Cov}[x_\alpha(t),x_\beta(t')]=\delta_{\alpha\beta}\sigma_0(t,t')$, where $\alpha,\beta$ label the different spatial coordinates. This is due to the fact that $\text{Cov}[x_\alpha(t),x_\beta(t')]$ is, at fixed $t,t'$, an isotropic rank-2 tensor and is therefore proportional to $\delta_{\alpha\beta}$. We further assume that each coordinate $x_\alpha(t)$ is a Gaussian stochastic process that has long time stationary increments, so that $\sigma_0$ satisfies the scaling behavior (\ref{ScalingSigma0}) with the condition (\ref{BehaviorGSummary}). 
If $d=1$,  we consider the  first-passage problem to a fixed threshold, for which the persistence $S(t)$, i.e. the probability of not having crossed this threshold, decays as $S(t)\propto 1/t^\theta$. The natural generalization of this definition to $d\ge1$ is to consider the first-passage problem to a target region (which is reduced to a point for $d=1$); here we assume that the process is \textit{compact} ($dH<1$), so that the target can be taken as punctual without any ambiguity. We also assume the process is \textit{non-smooth}. Our starting point is a generalized form of the renewal equation
\begin{align}
p(\ve[0],t)=\int_0^td\tau \ f(\tau) p(\ve[0],t\vert \mathrm{FPT}=\tau) \label{Ren1},
\end{align}
where $p(\ve[x],t)$ is the probability density to observe the position $\ve[x]$ (in $d$ dimensions) at time $t$, starting from the initial conditions at $t=0$, and $f(t)=-\partial_t S(t)$ is the probability density that the FPT is equal to $t$. Here we have used the hypothesis that the process is rough, and that it is compact, so that a punctual target can be found by the random walker. The above equation is simply obtained by partitioning the event of being inside the target at time $t$ over the value of the first-passage time. 

\begin{figure}[h!]
\includegraphics[width=12cm]{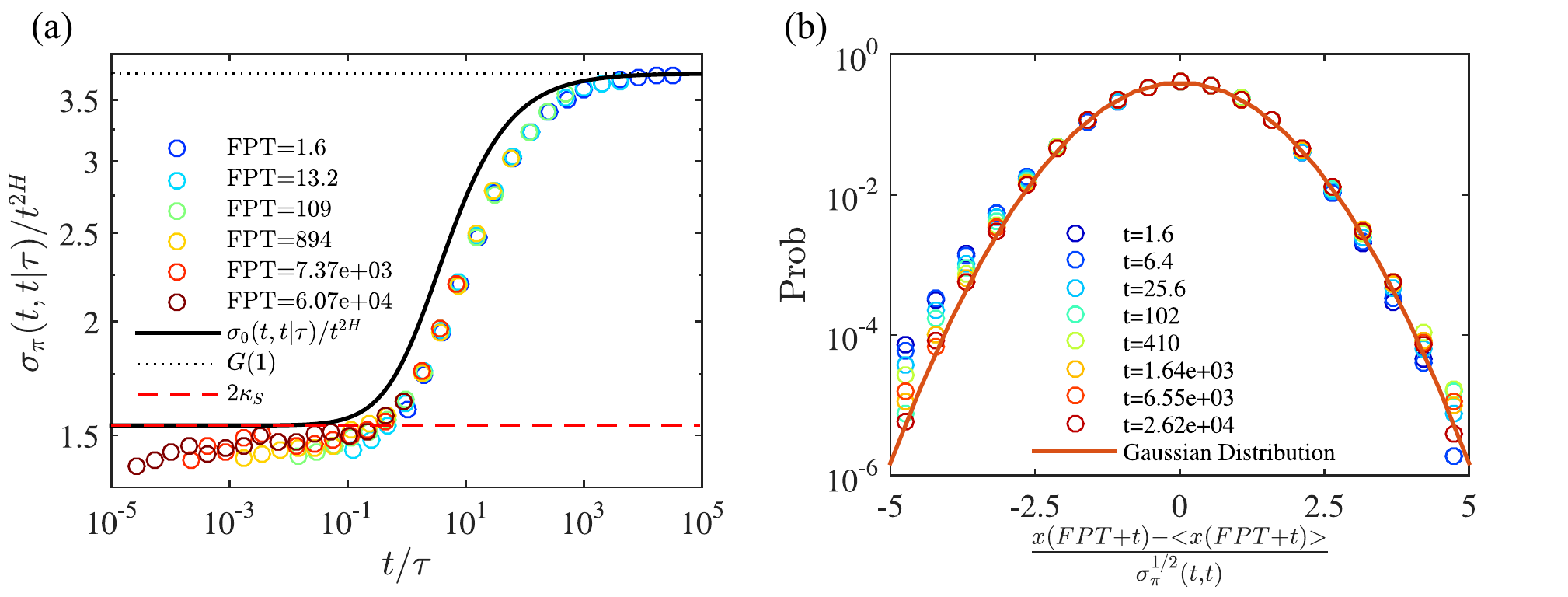} 
\caption{Results obtained for type I process, for $H=3/8$, $T=10$ and $d=1$. For these parameters, we have simulated numerically Eq.~(\ref{DynInterface}), for which $x(t)=h(s=0,t)$ starts at $x_0=1$. We have recorded the statistics of the trajectories after the first-passage to $x=0$, and identified 
the conditional variance of $x(t+\mathrm{FPT})$ conditioned on the value of the $\mathrm{FPT}=\tau$. This quantity is represented in (a) in rescaled variables; if the Gaussian hypothesis holds then the function appearing $B$ in Eq.~(\ref{Hyp1}) is $B(y)=1/\sqrt{2\pi\sigma_\pi(1,1\vert y)}$, hence these data support the fact that $B(y)$ is finite for $y\to\infty$ and $y\to0$. The black line represents the ansatz $B$ in Eq.~(\ref{DefAnsatz}). In (b) we have represented the normalized distribution of  $x(t+\text{FPT})$ for various values of $t$, small deviations from Gaussian behavior can be observed at small $t$ and disappear at larger values.}
\label{FigInterf1D}
\end{figure}

We define $p_\pi(x,t)$ as the probability density to observe $\ve[x]$ at a time $t$ after the FPT:
\begin{align}
p_\pi(\ve[x],t)\equiv \int_0^\infty d\tau \ p(\ve[x],t+\tau \vert \mathrm{FPT}=\tau)f(\tau). \label{pPi}
\end{align}
Next, we consider  a  fixed time $T$, and integrate the renewal equation (\ref{Ren1}) between $t=0$ and $t=T$:
\begin{align}
\int_0^Tdt\ p(\ve[0],t)=\int_0^Tdt\int_0^t d\tau f(\tau) p(\ve[0],t\vert \mathrm{FPT}=\tau)=\int_0^T d\tau \int_\tau^T dt f(\tau) p(\ve[0],t\vert \mathrm{FPT}=\tau),
\end{align}
where we have simply inverted the order of integration in the second equality. Setting $t=u+\tau$, we obtain
\begin{align}
\int_0^Tdt\ p(\ve[0],t)=\int_0^T d\tau \int_0^{T-\tau} du f(\tau) p(\ve[0],\tau+u\vert \mathrm{FPT}=\tau)
=\int_0^Tdu \int_0^{T - u} d\tau f(\tau) p(\ve[0],\tau+u\vert \mathrm{FPT}=\tau),\label{942M}
\end{align}
where, again the order of integration has been inverted in the second equality. 
Now, we integrate  Eq.~(\ref{pPi}) over $t$ between $t=0$ and $t=T$ and use Eq.~(\ref{942M}) to obtain
\begin{align}
\int_0^Tdt\ [p_\pi(\ve[0],t)-p(\ve[0],t)]=\int_0^Tdu \int_{T - u}^\infty d\tau f(\tau) p(\ve[0],\tau+u\vert \mathrm{FPT}=\tau). \label{032}
\end{align}
We will  evaluate   both sides of this equation for large $T$. To evaluate the right hand side (rhs) term, we assume  that 
\begin{align}
p(\ve[0],\tau+u\vert\mathrm{FPT}=\tau)\sim B(\tau/u)/u^{dH} \hspace{1cm}(u\to\infty,\tau\to\infty) \label{Hyp1},
\end{align}
where $B$ is a scaling function that is assumed to be finite and admit finite limits for small and large arguments. The above hypothesis is reasonable, because in the future $u$ of the FPT, one expects that the probability distribution of $\ve[x]$ will be some function whose extension is proportional $L\sim u^{H}$, so that the probability to be at $\ve[x]=\ve[0]$ is $\sim1 /L^d=1/u^{dH}$. The  prefactor that can depend on $u/\tau$, and the hypothesis that $B$ is finite is well supported by our simulations (see e.g. Fig.~\ref{FigInterf1D}).  
Now, by definition of $\theta$, we have for large times 
\begin{align}
f(t)\underset{t\to\infty}{\simeq} f_0/t^{\theta+1},
\end{align}
where $f_0$ is a prefactor. The term on the right hand side of Eq.~(\ref{032}) can be evaluated in the large $T$ limit by setting $\tilde{u}=u/T$ and $\tilde{\tau}=\tau/T$ and taking the large $T$ limit in the resulting expression, leading to
\begin{align}
 \int_0^Tdu \int_{T - u}^\infty d\tau f(\tau) p(\ve[0],\tau+u\vert \mathrm{FPT}=\tau) & =  T^2 \int_0^1d\tilde{u} \int_{1 - \tilde{u}}^\infty d\tilde{\tau} f(T\tilde{\tau}) p(\ve[0],T(\tilde{\tau}+\tilde{u})\vert \mathrm{FPT}=T\tilde{\tau}) \\
 & \underset{T\to\infty}{\simeq} T^{1-\theta-dH} \int_0^1d\tilde u \ Q(\tilde{u}) \label{DefQ},
 \end{align}
where $Q$ is given by 
 \begin{align}
 Q(\tilde{u})&=\int_{1-\tilde u}^\infty d\tilde \tau \frac{ f_0  B(\tilde{\tau}/\tilde{u}) }{\tilde\tau^{1+\theta}\tilde{u}^{dH}}=\frac{f_0}{\tilde{u}^{\theta+dH}} \int_{(1-\tilde u)/\tilde{u}}^\infty d\xi   \frac{   B(\xi) }{\xi^{1+\theta} }.
 \end{align}
The existence of $Q$ is guaranteed by the fact that  $\theta>0$. Its asymptotics  read
\begin{align}
Q(\tilde{u}\to0)\simeq \frac{f_0}{\tilde{u}^{\theta+dH}}    \frac{   B(\infty) }{ \theta \tilde{u}^{-\theta} } \simeq  \frac{ f_0  B(\infty) }{ \theta \ \tilde{u}^{dH } }, \hspace{2cm}
Q(\tilde{u}\to1) \simeq f_0    \frac{   B(0) }{ \theta \ (1-\tilde{u})^{\theta} }.
\end{align}
The above relations suffice to establish that $\int_0^1 dx Q(x)$ exists if $\theta<1$ and $dH<1$. The condition $dH<1$ was already assumed (compact process), and from now on we also assume $\theta<1$ (meaning that the mean first-passage time is infinite).  In these conditions, we deduce the scaling 
\begin{align}
\int_0^Tdu \int_{T - u}^\infty d\tau f(\tau) p(\ve[0],\tau+u\vert \mathrm{FPT}=\tau)  \propto T^{1-\theta-dH}. \label{0P594323}
 \end{align}
We now   evaluate the term in the left hand side of Eq.~(\ref{032}). At leading order, dimensional analysis indicate that
\begin{align}
p(\ve[0],t) \underset{t\to\infty}{\sim} \frac{K_0}{t^{dH}}, \hspace{1cm}  p(\ve[0],t) \underset{t\to\infty}{\sim} \frac{K_\pi}{t^{dH}}, 
\end{align}
where $K_0$ and $K_\pi$ are constants. Since $dH<1$, if $K_0\neq K_\pi$, we have at leading order
\begin{align}
\int_0^Tdt\ [p_\pi(\ve[0],t)-p(\ve[0],t)]\underset{T\to\infty}\sim \frac{(K_\pi-K_0)}{1-dH}T^{1-dH}.
\end{align}
This behavior must be  matched with that of the rhs of (\ref{032}), which behaves as $T^{1-dH-\theta}$ from Eq.~(\ref{0P594323}). Since $\theta>0$, we obtain a contradiction, which means that at long times, $p(\ve[0],t)$ and $p_\pi(\ve[0],t)$ are equal at leading order:
\begin{align}
K_0=K_\pi. \label{K0KPI}
\end{align}
We thus postulate an exponent $\alpha>dH$ for the next-to-leading order correction, such that
\begin{align}
p_\pi(\ve[0],t)-p(\ve[0],t)\simeq \frac{K_1}{t^{\alpha}}+... \hspace{1cm}(t\to\infty).
\end{align}
In order to find the value of $\alpha$, we have to discuss according to the value of $\theta$. If $\theta<1-dH$, we see from Eq.~(\ref{0P594323}) that the rhs  of Eq.~(\ref{032}) diverges with $T$ for large $T$. This means that $\alpha<1$, so that the left hand side term of Eq.~(\ref{032}) also diverges with $T$, as $T^{1-\alpha}$. Equating the divergences leads to
\begin{align}
\alpha=dH+\theta  \label{EqAlpha}.
\end{align}
In the opposite case $\theta>1-dH$, we see from Eq.~(\ref{0P594323}) that the rhs  of Eq.~(\ref{032}) tends to $0$ with $T$ for large $T$. This means that the following condition must hold exactly: 
\begin{align}
\int_0^{\infty}dt\ [p_\pi(\ve[0],t)-p(\ve[0],t)]=0\hspace{1cm}(\text{if } 1>\theta>1-dH). \label{9R432}
\end{align}
Then, using the trick $\int_0^T(...)=\int_0^\infty(...)-\int_T^\infty(...)$, we find 
\begin{align}
-\int_T^\infty [p_\pi(\ve[0],t)-p(\ve[0],t)]\propto T^{1-dH-\theta},
\end{align}
which leads to $\alpha=dH+\theta $ again.  
To proceed further, we define the covariance in the future of the first-passage time, 
\begin{align}
\sigma_\pi(t,t')=\text{Cov}[x(\mathrm{FPT}+t),x(\mathrm{FPT}+t')],
\end{align}
and we define an exponent $\beta$   characterizing the divergence of the mean square displacement (MSD) after the FPT as
\begin{align}
\sigma_\pi(t,t)-\sigma_0(t,t)  \simeq  c_1\ t^{\beta} \hspace{1cm}(t\rightarrow\infty) \label{042245}.
\end{align}
If we assume that the stochastic process in the future of the FPT is Gaussian (which is well supported by our simulations, see e.g. Fig.~\ref{FigInterf1D}), we obtain $p_\pi(\ve[0],t)\propto \sigma_\pi(t,t)^{-d/2} $. This leads to $\alpha=(d+2)H-\beta$, and thus  
\begin{align}
\beta=2H-\theta \label{ValueBeta}
\end{align}
Inserting Eq.~(\ref{ValueBeta}) into Eq.~(\ref{042245}), we obtain Eq.~(3) in the main text. In summary, here we have identified \textit{a link between the persistence exponent $\theta$ and the exponent $\beta$ which characterizes the MSD of the trajectories in the future of the FPT at long times}. Let us remind our hypotheses: (i) $dH<1$ (compact process), (ii) $\theta<1$ (infinite mean FPT), (iii) that the scaling function $B$ characterizing the propagator in the future of the FPT is finite for small and large arguments, and (iv) the trajectories in the future of the FPT display Gaussian statistics. The validity of (iii) and (iv) can be appreciated on Fig.~\ref{FigInterf1D} in one example.

\subsection{Self-consistent equations for the covariance of trajectories in the future of the FPT} 
\label{SectionEqCov}
We now write the generalization of the renewal equation  (\ref{Ren1}) for the probability to observe $\ve[x]=0$ at $t$ and the sequence of positions $\ve[x]_1,\ve[x]_2$ at times $t+t_1,t+t_2$, where  $t_1,t_2$ are fixed times:
\begin{align}
p(\ve[0],t;\ve[x]_1,t+t_1;\ve[x]_2,t+t_2)=\int_0^td\tau f(\tau) \ p(\ve[0],t;\ve[x]_1,t+t_1;\ve[x]_2,t+t_2 \vert \mathrm{FPT}=\tau).
\end{align}
We also define the probability density to observe the sequence $(\ve[0]; \ve[x]_1;\ve[x]_2)$ at respective times $t,t+t_1,t+t_2$ in the future of the FPT, averaged over the FPT distribution:
\begin{align}
p_\pi(\ve[0],t; \ve[x]_1,t+t_1; \ve[x]_2,t+t_2)=\int_0^\infty d\tau f(\tau) p(\ve[0],t+\tau; \ve[x]_1,t+\tau +t_1;\ve[x]_2,t+\tau +t_2 \vert \mathrm{FPT}=\tau).
\end{align}
Repeating the steps that had led to Eq.~(\ref{032}), we obtain
\begin{align}
\int_0^Tdt [p_\pi(\ve[0],t;\ve[x]_1,t+t_1;\ve[x]_2,t+t_2) &-p(\ve[0],t; \ve[x]_1,t+t_1;\ve[x]_2,t+t_2) ]=\nonumber\\
&\int_0^Tdu \int_{T - u}^\infty d\tau f(\tau) p(\ve[0],\tau+u;\ve[x]_1,u+\tau +t_1;\ve[x]_2,u+\tau +t_2\vert \mathrm{FPT}=\tau) \label{032G}.
\end{align}
Here we assume, without loss of generality, that the process is centered (zero average) ; we stress that this has no consequence on the value of $\theta$ which does not depend on microscopic details. We  multiply the above equation by $x_1 x_2$ and integrate over all positions to obtain the equation for the covariances: 
\begin{align}
\int_0^Tdt\ [&p_\pi(\ve[0],t)\sigma_\pi^*(t_1,t_2\vert t)   -p(\ve[0],t) \sigma^*(t_1,t_2\vert t)] = \int_0^Tdu \int_{T - u}^\infty d\tau f(\tau) p(\ve[0],\tau+u\vert \mathrm{FPT}=\tau) \Sigma(t_1,t_2,u,\tau)   \equiv C(t_1,t_2,T)  \label{032GCov}.
\end{align}
Here, the term $C$ is defined by this equation, and we have used the following notations for conditional covariances:
\begin{align}
&\sigma^*(t_1,t_2\vert t)= \mathrm{Cov}(x(t+t_1),x(t+t_2) \vert x(t)=0),\\
&\sigma_\pi^*(t_1,t_2\vert t)=   \mathrm{Cov}(x(\mathrm{FPT}+t+t_1),x(\mathrm{FPT}+t+t_2)\vert x(\mathrm{FPT}+t)=0),\\ 
&\Sigma(t_1,t_2,u,\tau)=\text{Cov}(x(u+\tau+t_1),x(u+\tau+t_2)\vert x(u+\tau)=0, \text{FPT}=\tau ).
\end{align}
Now, the main problem consists in evaluating $\Sigma$, which is difficult since it depends it-self on the value of the first-passage time. Nevertheless, we argue below that it is reasonable to assume that, when a long time has ocured after the first-passage, this conditional covariance becomes equal to the long-time stationary covariance:
\begin{align}
\Sigma(t_1,t_2,u,\tau) \sim \sigma_s(t_1,t_2)  \hspace{2cm}(\tau,u\to\infty). \label{HypSIGMA}
\end{align}
Here, we consider the above equality as an hypothesis, whose validity will be discussed later (see Section \ref{SectionValidity}). If Eq.~(\ref{HypSIGMA}) holds, then the term $C$ can be evaluated in the limit $T\to\infty$ as 
\begin{align}
C(t_1,t_2,T) = T^2 \int_0^1d\tilde{u} \int_{1 - \tilde{u}}^\infty d\tilde{\tau} f(T\tilde{\tau}) p(\ve[0],T[\tilde{\tau}+\tilde{u}\vert \mathrm{FPT}=T\tilde{\tau}) \Sigma(t_1,t_2,\tilde{u}T,\tilde{\tau}T)\nonumber \\
\underset{T\to\infty}{\sim}
  T^{1-\theta-dH} \int_0^1d\tilde u \int_{1-\tilde u}^\infty d\tilde \tau \frac{ f_0  B(\tilde{\tau}/\tilde{u}) }{\tilde\tau^{1+\theta}\tilde{u}^{dH}}\sigma_s(t_1,t_2)\label{EvalC},
\end{align}
with the same notations as in Eq.~(\ref{DefQ}). What is important here is the scaling with $T$. We see that if $\theta>1-dH$, $C$ tends to zero in the large $T$ limit, this suggests that in the large $T$ limit, Eq.~(\ref{032GCov}) becomes 
\begin{align}
\int_0^\infty dt \ [p_\pi(\ve[0],t)\sigma_\pi^*(t_1,t_2\vert t) - p(\ve[0],t)\sigma^*(t_1,t_2\vert t)]= 0.
\end{align}
This equation can be written explicitly as a function of the covariances $\sigma_0$ and $\sigma_\pi$:
\begin{align}
\mathcal{H}(\tau,\tau')\equiv\int_0^\infty dt\Bigg\{ \frac{\left[\sigma_\pi(t+\tau,t+\tau')-\frac{\sigma_\pi(t+\tau,t)\sigma_\pi(t+\tau',t)}{\sigma_\pi(t,t)}     \right]}{\sigma_\pi(t,t)^{d/2}} 
- \frac{\left[ \sigma_0(t+\tau,t+\tau')-\frac{\sigma_0(t+\tau,t)\sigma_0(t+\tau',t)}{\sigma_0(t,t)} \right]}{\sigma_0(t,t)^{d/2}}
\Bigg\}=0  \label{EqCovLargeTheta}.
 \end{align}
  This equation defines the covariance matrix for $\theta>1-dH$. However, it cannot be true for $\theta<1-dH$ since the above integral does not exist. Let us consider now the case $\theta<1-dH$, for which we see that $C$ diverges for large $T$ [Eq.~(\ref{EvalC})]. Our approach consists in combining Eqs.~(\ref{032GCov}) and Eq~(\ref{032}) [multiplied by $\sigma_s(t_1,t_2)$] to obtain 
 \begin{align}
\int_0^Tdt \{p_\pi(\ve[0],t)[\sigma_\pi^*(t_1,t_2,t)-&\sigma_s(t_1,t_2)] - p(\ve[0],t)[\sigma^*(t_1,t_2\vert t)-\sigma_s(t_1,t_2)]\} \nonumber \\
 = &\int_0^Tdu \int_{T - u}^\infty d\tau f(\tau) p(\ve[0],\tau+u\vert \mathrm{FPT}=\tau) [\Sigma(t_1,t_2,u,\tau)-\sigma_s(t_1,t_2)] \equiv R(t_1,t_2,T)     \label{9531}  ,
\end{align}
where $R$ is defined as the rhs of this equation. We note that $R$ is no longer of order $T^{1-dH-\theta}$, this problematic term has been removed. Assuming that $R$ vanishes for large $T$ (this will be  discussed in Section \ref{SectionValidity}), we obtain:
\begin{align}
\mathcal{H}(\tau,\tau')\equiv\int_0^\infty dt\Bigg\{ \frac{1
}{\sigma_\pi(t,t)^{d/2}}\Bigg[&\sigma_\pi(t+\tau,t+\tau')-\frac{\sigma_\pi(t+\tau,t)\sigma_\pi(t+\tau',t)}{\sigma_\pi(t,t)}  -  \sigma_s( \tau,\tau') \Bigg]\nonumber\\
- \frac{1}{\sigma_0(t,t)^{d/2}}\Bigg[&\sigma_0(t+\tau,t+\tau')-\frac{\sigma_0(t+\tau,t)\sigma_0(t+\tau',t)}{\sigma_0(t,t)}-\sigma_s( \tau, \tau') \Bigg]
\Bigg\}=0 \label{CorrEqSigma}.
 \end{align}
This equation is valid for $\theta<1-dH$; in fact it is also correct for $\theta>1-dH$, as can be seen by combining  Eqs.~(\ref{EqCovLargeTheta}) and   (\ref{9R432}). For $\theta=1-dH$ and $d=1$,  
we recover the equation for processes with stationary increments \cite{guerin2016mean}.  In summary, here we have obtained self-consistent equations [Eqs.~(\ref{EqCovLargeTheta}) and (\ref{CorrEqSigma})] for the covariance in the future of the first-passage time for non-stationary Gaussian processes. 

\subsection{Equation for the large time behavior of the trajectories after the first-passage [Derivation of Eq.~(4)] }

Up to now, the equations for $\sigma_\pi(t,t')$ involve all time scales of the process, including microscopic ones. Here, our goal is to isolate the contribution of large times. First, we note that it is natural to assume that $\sigma_\pi(t,t')\sim t^{2H}G_\pi(t/t')$ for large $t,t'$. We see that $\mathcal{H}(\tau,\tau')$ in  Eqs.~(\ref{EqCovLargeTheta}) and (\ref{CorrEqSigma}) is proportional to $\tau^{1-(d-2)H}$ multiplied by a scaling function of $\tau/\tau'$, and the fact that this term must vanish imposes that $G_\pi(x)=G(x)$ for all $x$. This equality generalizes the result (\ref{K0KPI}). Therefore, we look at large time corrections for $\sigma_\pi$, of the type
\begin{align}
\sigma_\pi(t,t')\simeq \sigma_0(t,t')+  \rho(t,t')+...\hspace{1cm}\rho(t,t')=  t^{2H-\theta}z(t/t'), \label{LongTimeCorr}
\end{align}
with $z$ a scaling function. In the following, we look for an equation that defines the function $z(x)$. 

First, we consider the case $\theta<1-dH$,  for which  $\sigma_\pi$ satisfies Eq.~(\ref{CorrEqSigma}), which we write for $\tau=Tv$ and $\tau'=Tv'$:
\begin{align}
\mathcal{H}(Tv,Tv')=  \int_0^\infty du \Bigg\{& \frac{T
}{\sigma_\pi(Tu,Tu)^{d/2}}\left[\sigma_\pi(T(u+v),T(u+v'))-\frac{\sigma_\pi(T(u+v),Tu)\sigma_\pi(T(u+v'),Tu)}{\sigma_\pi(Tu,Tu)}  -  \sigma_s( Tv,Tv') \right]\nonumber\\
- \frac{T}{\sigma_0(Tu,Tu)^{d/2}}&\left[\sigma_0(T(u+v),T(u+v'))-\frac{\sigma_0(T(u+v),Tu)\sigma_0(T(u+v'),Tu)}{\sigma_0(Tu,Tu)}-\sigma_s( Tv,Tv') \right]
\Bigg\}=0,
 \end{align}
where we have set $t=Tu$. Using (\ref{LongTimeCorr}), the evaluation of all terms when $T\to\infty$ leads to  
\begin{align}
\mathcal{H}&(Tv,Tv') \underset{T\to\infty}{\sim}  
 T^{1-dH+2H-\theta}\int_0^\infty \frac{du}{\sigma(u,u)^{d/2}}\Bigg\{   \rho(u+v,u+v')-\frac{\sigma(u+v,u)\rho(u+v',u)}{\sigma(u,u)}-\frac{\sigma(u+v',u)\rho(u+v,u)}{\sigma(u,u)}\nonumber\\
&+\rho(u,u)\frac{\sigma(u+v,u)\sigma(u+v',u)}{\sigma(u,u)^2}    - \frac{d\ \rho(u,u)}{2\ \sigma(u,u)}\left[\sigma(u+v,u+v')-\frac{\sigma(u+v,u)\sigma(u+v',u)}{\sigma(u,u)} -\sigma_s(v,v')\right]\Bigg\}=0. \label{09542}
\end{align}
It can be checked that this integral exists for $\theta<1-dH$. In particular, the integrand of the above expression for small $u$   is $\sim [\sigma(v,v')-\sigma_s(v,v')]/u^{dH+\theta} $, which is integrable since we have $\theta<1-dH$. Since $\mathcal{H}$ must vanish at all orders, equating the above expression to zero gives Eq.~(4) of the main text, that defines $\rho(v,v')$ for $\theta<1-dH$.  

If $\theta>1-dH$, we follow the same steps as for $\theta<1-dH$, but we start from Eq.~(\ref{EqCovLargeTheta}) [instead of (\ref{CorrEqSigma})]. We obtain  
\begin{align}
\mathcal{H}&(Tv,Tv')\underset{T\to\infty}{\sim}  
 T^{1-dH+2H-\theta}\int_0^\infty \frac{du}{\sigma(u,u)^{d/2}}\Bigg\{   \rho(u+v,u+v')-\frac{\sigma(u+v,u)\rho(u+v',u)}{\sigma(u,u)}-\frac{\sigma(u+v',u)\rho(u+v,u)}{\sigma(u,u)}\nonumber\\
&+\rho(u,u)\frac{\sigma(u+v,u)\sigma(u+v',u)}{\sigma(u,u)^2}    - \frac{d\ \rho(u,u)}{2\ \sigma(u,u)}\left[\sigma(u+v,u+v')-\frac{\sigma(u+v,u)\sigma(u+v',u)}{\sigma(u,u)} -\sigma(v,v')\right]\Bigg\}=0 \label{HDEO04312}.
\end{align}
Note that the only difference with Eq.~(\ref{09542}) is the last term, where $\sigma$ appears in place of $\sigma_s$.  For $u\to\infty$, the integrand of the above expression is $\sim d[\sigma_s(v,v')-\sigma(v,v')]/(2u^{dH+\theta})$, which is integrable  since $\theta>1-dH$. Equating the above expression to zero leads to Eq.~(4) in the main text for $\theta>1-dH$.

\subsection{The equation for $z(x)$ [Derivation of Eq.~(6)]}

Due to scaling properties, the solutions of Eqs.~(\ref{09542}) and (\ref{HDEO04312}) can be found under the scaling form $\rho(v,v')=v^{2H-\theta}z(v/v')$, where the single-valued function $z(x)$ satisfies a linear equation which we determine now. Using the same arguments as in Section \ref{SectionIncrLongTimes}, we can show that, if the process in the future of the FPT has long-time stationary increments, and if $z(x)$ has a linear term in its expansion near $x=1$, then 
\begin{align}
z(x)\underset{x\to1}{\simeq} z(1)[1+\beta/2(1-x)] +...
\end{align}
Since it is reasonable to believe that the process in the future of the FPT admits stationary increments at long times, it is also reasonable to assume the above equality holds, and it is actually the case for the functions $z$ determined at perturbative order, see Eqs.~(\ref{zPertQuenched}),(\ref{zPastFBMPert}). This suggests to pose 
\begin{align}
w(x)=\frac{z(x)}{z(1)}-\left[1+\frac{\beta}{2} (1-x)\right] \label{DefW},
\end{align} 
so that $w(1)=0$. At this stage, we note that it is perfectly equivalent to look for $w(x)$ or for $z(x)$, and we will see later that many difficulties can be overcome by using $w(x)$ instead of $z(x)$.  The equation for $w(x)$ is obtained by inserting Eqs.~(\ref{LongTimeCorr}) and (\ref{DefW}) into Eqs.~(\ref{09542}) and  (\ref{HDEO04312}), in which we also set $v=1$ and $v'=1/y$, with $0<y<1$. Splitting the resulting integral into a  term proportional to $w(x)$ and a second member, we obtain
\begin{align}
\int_0^\infty \frac{du}{u^{dH}}\left\{ (u+1)^{\beta}w\left(\frac{u+1}{u+y^{-1}}\right) -\frac{u^{\beta}}{G(1)}\left[G\left(\frac{u}{u+1}\right)w\left(\frac{u}{u+y^{-1}}\right)+G\left(\frac{u}{u+y^{-1}}\right)w\left(\frac{u}{u+1}\right)\right]\right\} =  f(y) \label{431},
\end{align}
where the second member $f(y)$ reads
\begin{align}
f(y)=&\int_0^\infty du\ I_f(u,y),\\
I_f(u,y)&= \frac{(-1)}{u^{dH}}\Big\{
(u+1)^{\beta}\left(1+ \frac{\beta(1-y)}{2(uy+1)}\right) 
-\frac{u^{\beta}}{G(1)}\left[G\left(\frac{u}{u+1}\right)\left(1+ \frac{\beta }{2(uy+1)}\right)
+G\left(\frac{uy}{uy+1}\right)\left(1+ \frac{\beta}{2(u+1)}\right)\right]  \nonumber\\
& + \frac{ u^{\beta}}{G(1)^2}    G\left(\frac{u}{u+1/y}\right)G\left(\frac{u}{u+1}\right)\left(1+\frac{d}{2}\right) 
-\frac{d}{2u^\theta G(1)} \left[(u+1)^{2H}G\left(\frac{u+1}{u+1/y}\right)    -G_K(y)\right]\Big\}\label{Def_f},
\end{align}  
with the notation 
\begin{align}
G_K(y)=
\begin{cases}
G_s(y) & \text{if } \theta<1-d H, \\
G(y) & \text{if }   \theta>1-d H.
\end{cases}
\end{align}
Eq.~(\ref{431})  can be written under the simpler form
\begin{align}
\int_0^1 dx \ K(x,y)\ w(x)=f(y) \label{EqInt}
\end{align}
where the kernel $K(x,y)$ is identified by separating the integrals for the 3 terms in the left-hand side of Eq.~(\ref{431}) and performing a change of variable in each of them, so that $x$ corresponds to the argument of $w$ (i.e. $x=(u_1+1)/(u_1+1/y)$ in the first term, $x=u_2/(u_2+1/y)$ in the second one and $x=u_3/(u_3+1)$ in the third one). This leads to
\begin{align}
K(x,y)= H(x-y) \frac{du_1}{dx}  \frac{(u_1+1)^{2H-\theta}}{u_1^{dH}}-\frac{du_2}{dx}  \frac{ u_2^{2H-\theta}}{u_2^{dH}G(1)}G\left(\frac{u_2}{u_2+1}\right)-\frac{du_3}{dx}  \frac{u_3^{2H-\theta}}{u_3^{dH}G(1)}G\left(\frac{u_3}{u_3+1/y}\right),
\end{align}
where $H(x)$ is the Heaviside step function, and 
\begin{align}
u_1=\frac{x - y}{(1 - x) y},  \hspace{1cm} u_2=\frac{x }{(1 - x) y},  \hspace{1cm} u_3=\frac{x }{1 - x}.
\end{align}

\textbf{Technical note for the numerical evaluation  of $f(y)$. } Of note,  evaluating numerically $f(y)$ is not straightforward because the integrand $I_f(u,y)$ does not always converge rapidly to zero for large $u$ (depending on the value of $\theta,H,d...$) and a similar problem arises for small $u$.  
The strategy to evaluate  $f$ numerically consists in identifying analytically the behavior of $f$ for small/large arguments and using this knowledge in the numerical program. More precisely, we identify the first terms in the expansion of $I_f(u,y)$ as 
\begin{align}
I_f(y,u)\underset{u\to\infty}{\simeq}  \frac{a_1^+}{u^{\nu_1^+}} + \frac{a_2^+}{u^{\nu_2^+}} + \frac{a_3^+}{u^{\nu_3^+}} +... , \hspace{1cm} I_f(y,u)\underset{u\to0}{\simeq} \frac{a_1^-}{u^{\nu_1^-}} + \frac{a_2^-}{u^{\nu_2^-}} + \frac{a_3^-}{u^{\nu_3^-}} +... \end{align}
Then, $f(y)$ can be evaluated as follows:
\begin{align}
f(y)= \int_0^1du \left(I_f(y,u)-\sum_{i=1}^{n^-} \frac{a_i^-}{u^{\nu_i^-}}\right) +\int_1^\infty du \left(I_f(y,u)-\sum_{i=1}^{n^+} \frac{a_i^+}{u^{\nu_i^+}}\right) +\sum_{i=1}^{n^-} \frac{a_i^-}{1-\nu_i^-}+\sum_{i=1}^{n^+} \frac{a_i^+}{\nu_i^+-1} \label{Evalf},
\end{align} 
where $n_+,n^-$ are the number of terms for which the expansion of $I_f$ is analytically known, and where now the integrals converge rapidly. This procedure thus require to identify the first order terms of the expansion of $I_f(u,y)$. After some algebra, using the expansion (\ref{BehaviorGSummary}) for $G$, we obtain
\begin{align}
\begin{cases}
\nu_1^+=dH+\theta, \hspace{2cm}  &a_1^+= d[G_s(y)-G_K(y)]/[2G(1)] \nonumber\\
\nu_2^+=dH+\theta+2H, \hspace{2cm}   &a_2^+=- \kappa_s^2\left(1+d/2\right)/G(1)^2 y^{2H}\nonumber\\
\nu_3^+ = dH+\theta+1-2H, \hspace{2cm}  &a_3^+=0\nonumber\\
\nu_4^+ =dH+\theta+1,\hspace{2cm}  &a_4^+ =    \frac{\kappa_s}{G(1)} \left(\frac{1}{y^{2H}}+\frac{1}{y}\right) \left[\left(1+\frac{d}{2}\right)H-\frac{\beta}{2} \right]-  \frac{d}{G(1)}H\kappa_s\left(1/y-1\right)^{2H}\nonumber 
\end{cases} 
\end{align}
and, for small arguments:
\begin{align}
\begin{cases}
\nu_1^-=dH+\theta,\hspace{2cm} &a_1^-= d [G(y)-G_K(y)]/[2G(1)] \nonumber\\
\nu_2^-=dH,\hspace{2cm} &a_2^-=-\left(1+\beta(1-y)/2\right)\nonumber\\
\nu_3^-=dH+\theta-2H,\hspace{2cm} &a_3^-=   2 \frac{G(0)}{G(1)} (1+\beta/2) - \frac{G(0)^2}{G(1)^2}\left(1+\frac{d}{2}\right)\nonumber\\
\nu_4^-=dH-2H+\theta-2\mu, \hspace{2cm}& a_4^-=     - \frac{\alpha_1^2}{G(1)^2}\left(1+\frac{d}{2}\right)y^{\mu}\nonumber\\
\nu_5^-=dH-2H+\theta- \mu,\hspace{2cm} &a_5^-= \frac{\alpha_1}{G(1)}\left(1+\frac{\beta}{2}\right)(1+y^{\mu}) - \frac{G_0\alpha_1}{G(1)^2}(1+y^{\mu}) \left(1+\frac{d}{2}\right)  \nonumber
\end{cases}
\end{align}
where $G_0, \alpha_1,\mu$ characterize the behavior of $G$ near zero: 
\begin{align}
G(x)\underset{x\to0}{\simeq} G_0 + \alpha_1 x^{\mu} +..., \hspace{3cm}
\mu=
\begin{cases}
1-2H   & (\text{type I})\\
1/2-H   & (\text{type II})\\
\end{cases}\nonumber
\end{align}

\subsection{Behavior of $z$ for small arguments and the criterion to select the correct value of $\theta$} 
\label{SelectTheta}
Numerical integration of $z(x)$ tends to show the presence of divergences for small values of $x$. Here, we give an argument to identify a  possible exponent $\alpha$ such that 
\begin{align}
z(x) \underset{x\to0}{\sim} \frac{1}{x^\alpha}\label{SMallz}.
\end{align}
Let us look for non-trivial values of $\alpha$, i.e. which are not contained in the expansions of $G(x)$. If one looks at Eq.~(\ref{431}) for large $y^{-1}$, we will find terms $\propto y^{-\alpha}$ that cannot be compensated by any other term if $\alpha$ is not trivial. The coefficient for this term must then vanish, leading to
\begin{align}
\int_0^\infty  \frac{dt}{t^{dH}} \left\{ \frac{(t+\tau)^{2H-\theta}}{(t+\tau)^\alpha} -\frac{t^{2H-\theta}}{t^\alpha }\times\frac{G(t/(t+\tau))}{G(1)} \right\}y^{-\alpha} + ... =0 .
\end{align}
The values of $\alpha$ that are authorized are those for which the prefactor of $y^{-\alpha}$ in the above equation vanish, this leads to 
\begin{align}
\alpha=\nu+2H-\theta,  \label{EqAlphaNu}
\end{align}
where $\nu$ is an exponent that does not depend on $\theta$ and satisfies the equation
\begin{align}
\int_0^\infty  \frac{dt}{t^{dH}} \left\{ \frac{1}{(t+1)^\nu} -\frac{1}{t^\nu }\times\frac{G(t/(t+1))}{G(1)} \right\}=0. 
\end{align}
When $G=G_s$ and $d=1$, a solution is $\nu=1-2H$ \cite{guerin2016mean} and $\theta=1-H$ so that $\alpha=H$. 
This means that most solutions $z(x)$ will diverge for small arguments. We postulate that the criterion to select the correct value of $\theta$ is to look for solutions that do not present this divergence as $1/x^\alpha$. 

\textbf{Procedure for the numerical evaluation of $\theta$. } To evaluate $\theta$, we have proceeded as follows: for a given ``test'' value of $\theta$, and a given choice of mesh points $y_i$, we evaluated the numerical solution of Eq.~(\ref{EqInt}). We observed that the obtained values of $z(x)$ behaved as predicted by (\ref{SMallz}),(\ref{EqAlphaNu}) and defined a coefficient $C_\theta$ so that $z(x)\simeq C_\theta/x^{\nu+2H-\theta}$. Repeating the procedure for several ``attempt'' values of $\theta$ enables us to identify the value of $\theta$ for which $C_\theta=0$; we chose this value as the output of our algorithm. We controlled that the results do not depend on the choice of the mesh. 

\section{Discussion of the validity of our hypotheses}
\label{SectionValidity}

Here, we discuss the validity of our theory. The first issue with our approach comes from the selection criterium of the correct value of $\theta$ via the argument that $z(x)$ (or, equivalently, $w(x)$) does not diverge at small values of $x$. This criterium seems plausible at least when  $2H-\theta>0$ in Eq.~(\ref{LongTimeCorr}), since for this value $\rho(t,t')\sim t^{2H-\theta}$ diverges for large $t$. However, this argument is less plausible when $2H-\theta<0$. This is supported by Fig.~\ref{FigAllTheta}, where we have compared the theoretical predictions of $\theta$ to the simulations for all the models presented in the main text, it is clearly seen that most deviations between theory and simulations appear when $\theta-2H>0$. We argue that a minimal criterium for our theory to work is to require that the condition $2H-\theta>0$ is satisfied at least for the stationary case ($\sigma=\sigma_s$), for which $\theta=1-dH$. \textit{This argument leads to the conclusion that we should restrict ourselves to the case $ H>1/(d+2)$}. 

\begin{figure}[h!]
\includegraphics[width=7cm]{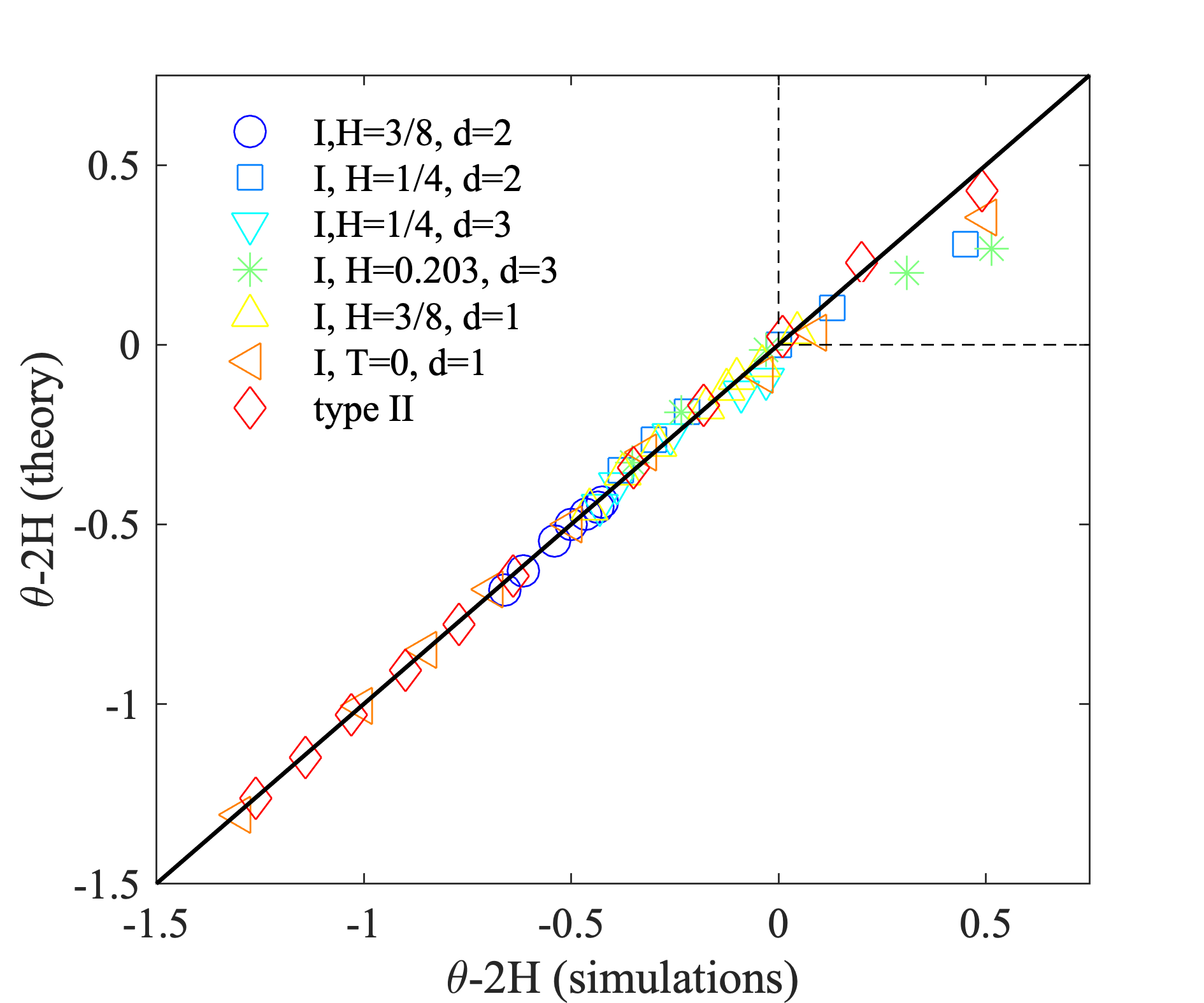} 
\caption{Comparison of theoretical values of the persistence exponent  versus simulations for all data shown in the main text.}
\label{FigAllTheta}
\end{figure}

Next, in Section \ref{SectionEqCov}, we have derived an equation that defines $\sigma_\pi$. However, the relation (\ref{HypSIGMA}) was assumed to hold, and we also assumed that the right-hand-side of Eq.~(\ref{9531}) tends to zero. Here we discuss these hypotheses. Since the evaluation of $\Sigma$ requires the knowledge of the covariance after the first-passage, a quantity which is unknown, it seems necessary to make some ansatz to evaluate it. We have tried two ansatz:
\begin{align}
\Sigma(t_1,t_2 ; u; \tau)\simeq \begin{cases}
\sigma_\pi^*(t_1,t_2\vert u) & \text{(Ansatz A: decoupling approx.)}\\
\text{Cov}[x(u+\tau+t_1),x(u+\tau+t_2)\vert x(u+\tau)=0;x(\tau)=0] & \text{(Ansatz B)}
\end{cases}\label{DefAnsatz}
\end{align}
In the decoupling approximation (Ansatz A), one assumes that the covariance in the future of the FPT does not depend on the value of this FPT, this can be seen as a ``mean-field'' approximation whose validity can be evaluated self-consistently. This approximation also reminds us the independent interval approximation to calculate the duration of intervals between successive zeros of a smooth Gaussian process, for which the duration of an interval is assumed to be independent on the duration of the previous interval. In turn, Ansatz B consists in replacing the condition that $x=0$ was reached for the first time at $\tau$ by the simpler condition that $x=0$ is observed at $\tau$ (not necessarily for the first time). This kind of approximation has been proposed before in the different context of polymer cyclization  \cite{Sokolov2003}. It is motivated by the fact that it leads to rather accurate predictions for the mean square displacement in the future of the first-passage time, see Fig.~\ref{FigInterf1D}. 

In Ansatz $A$, we need to remember that  $\sigma_\pi(t,t')\simeq \sigma_0(t,t')$ for large times, at leading order. It turns out that, in both ansatzs $A$ and $B$, the property $\Sigma\to\sigma_s$ [Eq.~(\ref{HypSIGMA})] holds, which is an argument in favor of the validity of our equations. Next, we note that 
\begin{align}
\Sigma(t_1,t_2 ; u; \tau)\underset{\tau\to\infty}{\simeq} \begin{cases}
\sigma_\pi(t_1,t_2\vert u) & \text{(Ansatz A),}\\
 \sigma_s(t_1,t_2\vert u) & \text{(Ansatz B).}\\
\end{cases}
\end{align}
This is enough to evaluate the rhs of Eq.~(\ref{9531}) as 
\begin{align}
R(t_1,t_2,T) \simeq \  &T \int_0^Tdu \int_{1 - u/T}^\infty d\tilde{\tau} f(T\tilde{\tau}) p(\ve[0],T\tilde{\tau}+u\vert \mathrm{FPT}=T\tilde{\tau}) [\Sigma(t_1,t_2,u,T\tilde{\tau})-\sigma_s(t_1,t_2)]\label{9431143} \\
=\ &T^{-\theta} \times \begin{cases}
\int_0^\infty du \int_{1}^\infty d\tilde{\tau} \frac{f_0}{\tilde{\tau}^{1+\theta}} \frac{[\sigma_\pi(t_1,t_2\vert u)-\sigma_s(t_1,t_2)]}{[2\pi \sigma_0(u,u)]^{d/2}} & \text{(Ansatz A)}\\
 \int_0^\infty du \int_{1}^\infty d\tilde{\tau} \frac{f_0}{\tilde{\tau}^{1+\theta}} \frac{[\sigma_s(t_1,t_2\vert u)-\sigma_s(t_1,t_2)]}{[2\pi \sigma_s(u,u)]^{d/2}} & \text{(Ansatz B)}
\end{cases}
\end{align}
In Eq.~(\ref{9431143}) we have set $\tau=T\tilde{\tau}$ and the next evaluation follows from taking the limit $T\to\infty$ of all terms. 
The integrals in the above equation converge as soon as $\theta>0$, $dH<1$ (condition obtained for small $u$), and $(d+2)H>1$ (condition obtained for large $u$, for which $\sigma_s(t_1,t_2\vert u)-\sigma_s(t_1,t_2)\propto1/u^{2H}$ from the property (\ref{ConvergenceToStatCov}), and the same holds for $\sigma_\pi$ when one assumes that it has also long-time stationary increments). This means that, if $(d+2)H>1$, the leading order of the lhs of Eq.~(\ref{9531}) is  $1/T^\theta$ for large $T$, which tends to zero, so that our equation for the covariance is correct (at least from the point of view of these investigations with our two ansatzs). Hence, we expect that our theory holds when $H>1/(d+2)$, this validity criterium is the same as the one found above. 

In order to investigate the behavior of $R$ for $H<1/(d+2)$, it is necessary to show that
\begin{align}
\Sigma(t_1,t_2,u,\tau)\underset{u,\tau\to\infty}{\simeq}
\begin{cases} \sigma_s(t_1,t_2)-\frac{\kappa_s^2 (t_1t_2)^{2H}}{G(1)u^{2H}} \hspace{1cm} &\text{(Ansatz A}),\\
\sigma_s(t_1,t_2) - \frac{\kappa_s^2 \ (t_1t_2)^{2H}}{ \sigma(u,u\vert\tau) } \hspace{1cm}&\text{(Ansatz B}),
\end{cases}
\end{align}
 where the result for Ansatz $B$ comes from general formulas on conditional covariances \cite{Eaton1983} and the property (\ref{ConvergenceToStatCov}). 
If $(d+2)H<1$, we thus evaluate the rhs of Eq.~(\ref{9531}) as 
\begin{align}
R(t_1,t_2,T)  &=     T^2 \int_{1 - \tilde{u}}^\infty d\tilde{\tau} f(T\tilde{\tau}) p(\ve[0],T\tilde{\tau}+T\tilde{u}\vert \mathrm{FPT}=T\tilde{\tau}) [\Sigma(t_1,t_2,T\tilde{u},T\tilde{\tau})-\sigma_s(t_1,t_2)]\\
&\simeq -\frac{T^{1-(d+2)H-\theta}}{(2\pi)^{d/2}}\kappa_s^2\times\begin{cases} 
\int_0^1d\tilde{u} \int_{1 - \tilde{u}}^\infty d\tilde{\tau} \frac{f_0}{\tilde{\tau}^{1+\theta}} \frac{1}{[ G(1)u]^{(d+2)H}} & \text{(Ansatz A)},\\
\int_0^1d\tilde{u} \int_{1 - \tilde{u}}^\infty d\tilde{\tau} \frac{f_0}{\tilde{\tau}^{1+\theta}} \frac{1}{[\sigma(u,u\vert\tau)]^{{d+2}H}} & (\text{Ansatz B}).
\end{cases}
\end{align}
These integrals exist when $0<\theta<1$ and   $(d+2)H<1$ (which is assumed here), so that in this case $R\propto T^{1-(d+2)H-\theta}$. This means that when $\theta<1-(d+2)H$, our equation for the covariance is wrong since the rhs of  (\ref{9531}) does not vanish for large $T$.  Since there is no reason why $\theta$ should be larger than $1-(d+2)H$, we consider that using our formalism for $(d+2)H<1$ is not advised.

\begin{figure}[h!]
\includegraphics[width=11cm]{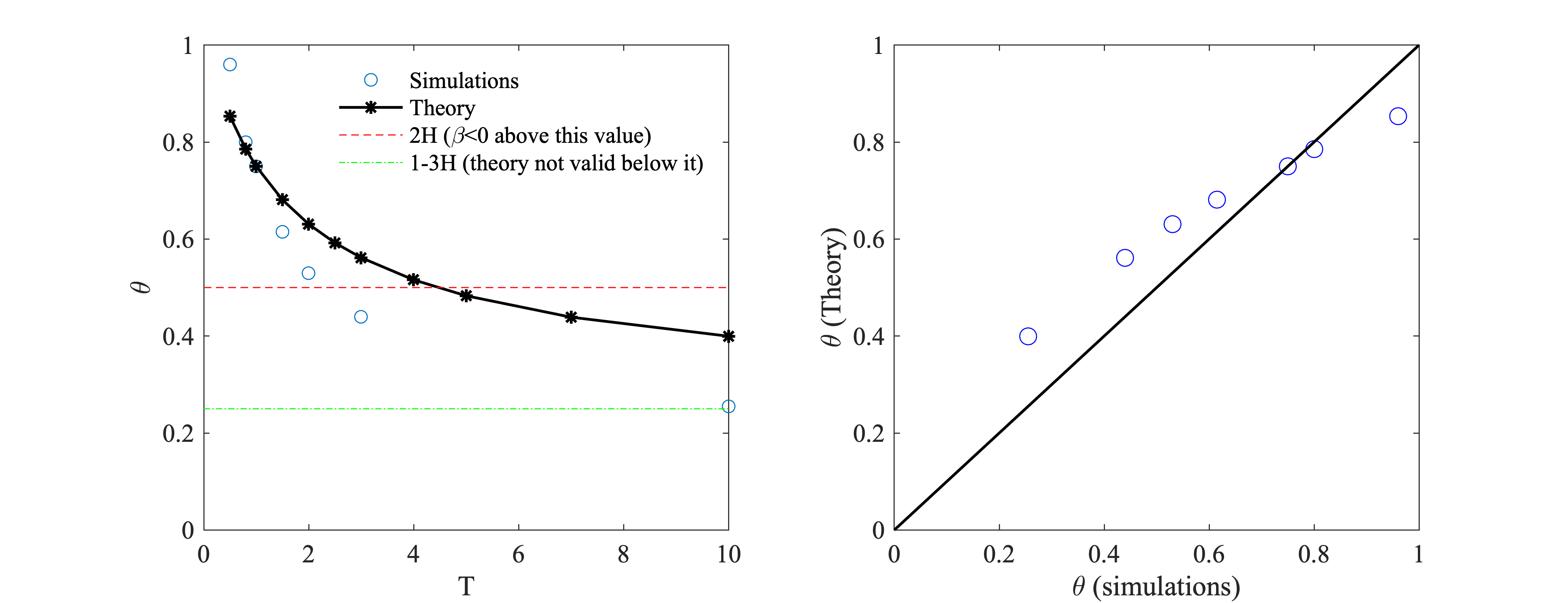} 
\caption{Results for $H=1/4$ and $d=1$ (which lies outside the validity criterium $H>1/(d+2)$ of our theory, see text).}
\label{FigRouse1D}
\end{figure}

In Fig.~\ref{FigRouse1D}, we present simulation results for a value of $H$ which does not satisfy this criterium ($H=1/4,d=1$). It is seen that the curve $\theta(T)$ is qualitatively correct, but there is a clear difference between the predicted value of $\theta$ and the measured one. However, from the arguments above, we are unsure of our criterium as soon as $\theta>2H=1/2$, and we know that our equation for the covariance is wrong for $\theta<1-3H=0.25$. This leaves a very small range of values of $\theta$ in which the theory could be expected to be correct (which does not even include the stationary state $T=1$) and this is probably the reason of the differences between theory and simulations. 

\section{Perturbative analysis of the formalism [Derivation of Eqs. (7),(8)]}
\label{PertSection}

Here, we perform an analysis at perturbative order of our equations in dimension $d=1$. We assume $H=1/2+\varepsilon$ with $\varepsilon$ small and we start directly with the equation (\ref{HDEO04312}) for $\rho(u,v)$. We assume $\theta>1-H$ (but we expect that the expansion in powers of $\varepsilon$ will be the same for $\theta>1-H$ or $\theta<1-H$):
  \begin{align}
 \int_0^\infty \frac{du}{u^H}\Bigg\{    \rho(u+v,u+v')-&\frac{\sigma(u+v,u)\rho(u+v',u)}{\sigma(u,u)}-\frac{\sigma(u+v',u)\rho(u+v,u)}{\sigma(u,u)}  +\rho(u,u)\frac{\sigma(u+v,u)\sigma(u+v',u)}{\sigma(u,u)^2} \Bigg\}   \nonumber\\
 &= \int_0^\infty \frac{du}{2u^{H+\theta}}   \left[\sigma(u+v,u+v')-\frac{\sigma(u+v,u)\sigma(u+v',u)}{\sigma(u,u)} -\sigma(v,v')\right]\equiv C(v,v'), \label{StartPertTheory}
\end{align}
where $C$ is defined by this equation. Following our formalism, we will look for symmetric solutions with $\rho(u,u)=u^{2H-\theta}$ and that our criterium to select the correct value of $\theta$ is that $\rho(u,v)$ does not diverge with $v$ as $v\to0$. Let us write the expansion of $\theta$ as:  
\begin{align}
\theta=1-H + \varepsilon \delta_1 +\varepsilon^2\delta_2+...,
\end{align}   
with $\delta_1\varepsilon>0$ to have $\theta>1-H$. Now, let us introduce the following notations  
\begin{align}
&\sigma_s(t,t')=\text{min}(t,t')+\varepsilon\sigma_{s,1}(t,t')+\varepsilon^2\sigma_{s,2}(t,t')+...
&\sigma(t,t')=\sigma_s(t,t')+\varepsilon\omega_1(t,t')+\varepsilon^2\omega_2(t,t')+...& \\
 &\rho(u,v)=\rho_0(u,v)+\varepsilon \rho_1(u,v)+\varepsilon^2 \rho_2(u,v)...&
\end{align}
Here, $\omega_i$ represents the  deviation of $\sigma$ with respect to $\sigma_s$ at order $i$. Note that the leading order term for $\rho$ is of order zero because of the normalization $\rho(u,u)=u^{2H-\theta}\simeq \sqrt{u}$ when $H\to1/2$. 
 
\subsection{First order}

To estimate the limit of $C$ for $\varepsilon\to0$, let us introduce a parameter $L$ (finite but large) and consider the calculation
\begin{align}
C(v,v')&= \int_0^L \frac{du}{2u^{H+\theta}}   \left[\sigma(u+v,u+v')-\frac{\sigma(u+v,u)\sigma(u+v',u)}{\sigma(u,u)} -\sigma(v,v')\right]+ \int_L^\infty \frac{du}{2u^{H+\theta}}   \left[\sigma_s(v,v') -\sigma(v,v')\right],\nonumber\\
 &= \int_0^L \frac{du}{2u^{H+\theta}}   \left[\sigma(u+v,u+v')-\frac{\sigma(u+v,u)\sigma(u+v',u)}{\sigma(u,u)} -\sigma(v,v')\right]+ \frac{L^{-\varepsilon \delta_1}}{2\varepsilon\delta_1} \left[\sigma_s(v,v') -\sigma(v,v')\right],\nonumber\\
& \underset{\varepsilon\to0}{\simeq} -  \frac{\omega_1(v,v')}{2\delta_1}.
\end{align}
We thus obtain 
  \begin{align}
 \int_0^\infty \frac{du}{u^{1/2}}\Bigg\{    \rho_0(u+v,u+v')- \rho_0(u+v',u) - \rho_0(u,u+v') +\rho_0(u,u) \Bigg\} 
 = - \frac{  \omega_1(v,v')}{2 \delta_1} \label{EqFirstOrder}.
\end{align}
Let us take the derivative with respect to $v$ and $v'$ of this equation:
\begin{align}
 \int_0^\infty \frac{du}{u^{1/2}}   \rho_0^{(1,1)}(u+v,u+v') 
 = - \frac{  \omega_1^{(1,1)}(v,v')}{2 \delta_1}, \label{EqIntPer}
\end{align}
with the notation $G^{(1,1)}=\partial_v\partial_{v'}G$ for all functions $G(v,v')$. The solution of this equation can be found by the method of superposition of solutions. Consider the simpler equation with a single complex exponential in the second member
\begin{align}
 \int_0^\infty \frac{du}{u^{1/2}}   \rho_0^{(1,1)}(u+v,u+v')  = e^{i(\omega v +\omega' v')}, 
\end{align}
for which an obvious solution is 
\begin{align}
  \rho_0^{(1,1)}(v,v')  = e^{i(\omega v +\omega' v')} \frac{1}{\int_0^\infty \frac{du}{u^{1/2}}e^{i(\omega+\omega')u}}.
\end{align}
Introducing the double Fourier transform  and its inverse for all functions $f(t,t')$ as
\begin{align}
\hat{f}(\omega,\omega')=\int_{-\infty}^\infty dt \int_{-\infty}^\infty dt' e^{-i(\omega t+\omega't')} f(t,t') , \hspace{1cm} f(t,t') =\frac{1}{4\pi^2}\int_{-\infty}^\infty d\omega \int_{-\infty}^\infty d\omega' e^{ i(\omega t+\omega't')} \hat{f}(\omega,\omega'),
\end{align}
we see that
\begin{align}
  \rho_0^{(1,1)}(v,v')  = -\frac{1}{8\pi^2\delta_1}\int_{-\infty}^\infty d\omega\int_{-\infty}^\infty d\omega' \ \widehat{\omega_1^{(1,1)}}(\omega,\omega') e^{i(\omega v +\omega' v')} \frac{1}{\int_0^\infty \frac{du}{u^{1/2}}e^{i(\omega+\omega')u}}
\end{align}
is a solution of (\ref{EqIntPer}) constructed by superposition of solutions.
Integrating with respect to $v,v'$ and using $\rho_1(v,0)=\rho_1(0,v')=0$,  we obtain
\begin{align}
&\rho_0(\xi,\xi')=\int_0^\xi dv \int_0^{\xi'} dv'   \rho_0^{(1,1)}(v,v') \nonumber\\
&= -\frac{1}{8\pi^2\delta_1}\int_{-\infty}^\infty d\omega\int_{-\infty}^\infty d\omega' \int_0^\infty dt \int_0^\infty dt' \omega_1^{(1,1)} (t,t') \frac{(e^{i \omega (\xi-t)}-e^{-i\omega t}) (e^{i \omega' (\xi'-t')}-e^{-i\omega' t'})   }{(-\omega \omega')} \frac{\sqrt{-i(\omega+\omega')}}{\sqrt{\pi}}\label{643342}.
\end{align}
We introduce a function $K_0(t,t')$ and its double Fourier transform  
\begin{align}
K_0(t,t')=\left(\frac{H(t'-t)}{\pi \sqrt{t}}+\frac{H(t-t')}{\pi \sqrt{t'}}\right)H(t)H(t'), \hspace{1cm} \widehat{K}_0(\omega,\omega')=-\frac{\sqrt{i(\omega+\omega')}}{\omega\omega'\sqrt{\pi}}\label{DefK0},
\end{align}
with $H$ the Heaviside step function. Therefore, if we change $\omega\to-\omega$ and $\omega'\to-\omega'$ in (\ref{643342}) and integrate over $\omega,\omega'$ before $t,t'$, we recognize the double inverse Fourier transform of $K$ and we obtain
\begin{align}
\rho_0(\xi,\xi') 
&= -\frac{1}{2 \delta_1} \int_0^\infty dt \int_0^\infty dt' \omega_1^{(1,1)} (t,t')  [K_0(t-\xi,t'-\xi')-K_0(t-\xi,t')-K_0(t,t'-\xi')+K_0(t, t')]. 
 \end{align}
Now, using (\ref{DefK0}) this equation can be integrated once  \cite{guerin2016mean}, leading to
\begin{align}
\rho_0(\xi,\xi')=\frac{1}{2\pi\delta_1} \int_0^\infty\frac{dt}{\sqrt{t}}[\omega_1^*(t+\xi,t+ \xi') -\omega_1^*(t+\xi,t ) -\omega_1^*(t ,t+ \xi') +\omega_1^*(t,t)]\label{GenSol1},
\end{align}
with $\omega_1^*$ is a symetrized first derivative of $\omega_1$, i.e. 
\begin{align}
\omega_1^*(t,t')=(\partial_t+\partial_{t'})\omega_1(t,t').
\end{align}
Eq.~(\ref{GenSol1}) is the solution of Eq.~(\ref{EqFirstOrder}). 
For the quenched fBm, we can calculate  
\begin{align}
\omega_1^*(t,t')=\lambda[2\ln(t+t')-\ln t -\ln t'] \label{omega1Star},
\end{align}
with $\lambda=1-T$. 
Inserting this value into Eq.~(\ref{GenSol1}) and performing the integral, we obtain 
\begin{align}
\rho_0(\xi,\xi')=  \frac{\lambda \sqrt{2}}{ \delta_1}  [\sqrt{\xi+\xi'}-\sqrt{\xi}-\sqrt{\xi'}]  \label{ExpRho1} .
\end{align}
Comparing  the above equation with the property   $\rho_0(\xi,\xi)=\sqrt{\xi}$ (which holds by the very definiton of $\rho_0$) enables us to identify $\delta_1$:
\begin{align}
\delta_1= - 2 \lambda (\sqrt{2}-1).
\end{align}
This is consistent with (and generalizes) the results of Ref \cite{Krug1997} for $\lambda=1$ and $\lambda=0$. 
It is also instructive to give the value of $z(x)$ at this order:
\begin{align}
z(x)=  -\frac{\sqrt{1+x}-\sqrt{1}-\sqrt{x}}{ (2-\sqrt{2})\sqrt{x}}.  \label{zPertQuenched}
\end{align}

\subsection{Second order (quenched fBm)}
At second order, we need the next-to-leading order of the expansion of the term  $C$ in Eq.~(\ref{StartPertTheory}):
\begin{align}
&C(v,v') =\int_0^L \frac{du}{2u^{1+\varepsilon\delta_1}}   \left[\sigma(u+v,u+v')-\frac{\sigma(u+v,u)\sigma(u+v',u)}{\sigma(u,u)} -\sigma(v,v')\right]+ \frac{L^{-\varepsilon \delta_1-\varepsilon^2\delta_2}}{2[\varepsilon\delta_1+\varepsilon^2\delta_2]} \left[\sigma_s(v,v') -\sigma(v,v')\right]\nonumber\\
& \simeq \varepsilon  \int_0^L \frac{du}{2u}   \left[\sigma_1(u+v,u+v')-\sigma_1(u+v,u)-\sigma_1(u,u+v')+\sigma_1(u,u)-\sigma_1(v,v') \right]+ \frac{1-\varepsilon\delta_1\ln L}{2\varepsilon \delta_1(1+\varepsilon\delta_2/\delta_1)} \left[-\varepsilon\omega_1-\varepsilon^2\omega_2\right] \nonumber\\
&\simeq - \frac{\omega_1(v,v')}{2\delta_1} +\varepsilon   C_2
\end{align}
with
\begin{align}
C_2 (v,v') =  \int_0^L \frac{du}{2u}  & \left[\omega_1(u+v,u+v')-\omega_1(u+v,u)-\omega_1(u,u+v')+\omega_1(u,u)-\omega_1(v,v') \right]  \nonumber\\
&+ \frac{\omega_1(v,v')\ln L }{2}+\omega_1(v,v')\frac{\delta_2}{2  \delta_1^2} - \frac{\omega_2(v,v')}{2\delta_1} 
\end{align}
Using $\ln L=\int_1^L du/u$ in the above expression, we see that $C_2$  is independent of $L$ and reads
\begin{align}
C_2 (v,v') = & \int_0^1 \frac{du}{2u}   \left[\omega_1(u+v,u+v')-\omega_1(u+v,u)-\omega_1(u,u+v')+\omega_1(u,u)-\omega_1(v,v') \right] +\nonumber\\
&\int_1^\infty \frac{du}{2u}   \left[\omega_1(u+v,u+v')-\omega_1(u+v,u)-\omega_1(u,u+v')+\omega_1(u,u) \right] 
+\omega_1(v,v')\frac{\delta_2}{2  \delta_1^2} - \frac{\omega_2(v,v')}{2\delta_1}. 
\end{align}
Expanding  Eq.~(\ref{StartPertTheory}) at second order, we find that the equation for $\rho_1$ is 
\begin{align}
& \int_0^\infty \frac{du}{u^{1/2}}\Bigg\{    \rho_1(u+v,u+v')-\rho_1(u+v,u)- \rho_1(u,u+v')+ \rho_1(u,u)\Bigg\}=A_1+A_2+C_2 \equiv A(v,v') \label{EqIntSecondOrder}\\
&A_1= \int_0^\infty \frac{du}{u^{3/2}}\Big\{[\sigma_1(u+v,u)-\sigma_1(u,u)][\rho_0(u+v',u)-\rho_0(u,u)] +[\sigma_1(u+v',u)-\sigma_1(u,u)][\rho_0(u+v,u)-\rho_0(u,u)] \Big\} \nonumber\\
&A_2=\int_0^\infty du \left(\frac{  \ln u}{u^{1/2}}\right) \Bigg\{\rho_0(u+v,u+v')-\rho_0(u+v,u)- \rho_0(u,u+v')+ \rho_0(u,u)\Bigg\} 
 \end{align}
Eq.~(\ref{EqIntSecondOrder}) is the same integral equation that appeared at first order, its solution with no divergence at the origin is 
\begin{align}
\rho_1(\xi,\xi)=-\frac{1}{\pi } \int_0^\infty\frac{dt}{\sqrt{t}}[A^*(t+\xi,t+ \xi)-2 A^*(t+\xi,t )  +A^*(t,t)] \label{rho1Quenched}.
\end{align}
Performing the integrations, we obtain
\begin{align}
&\rho_1(\xi,\xi)=\sqrt{\xi} [\alpha +\Omega \ln\xi], \hspace{1cm} \Omega=3 + 2 (-1 + \sqrt{2}) \lambda= 3-\delta_1\label{OmegaComplete}.\\
&\alpha=-4 + c + \frac{(1 + \sqrt{2}) \delta_2}{2 \lambda} + 
 \lambda (4 - 4 \sqrt{2} + b - 6 \ln2  + \sqrt{2} \ln 4) - 
 \sqrt{2}\ln 8 
\end{align}
where the numerical constants $b$ and $c$ are   defined by the integrals
\begin{align}
c=&-\frac{(2 + \sqrt{2}) }{\pi}\int_0^\infty \frac{d\tau}{\sqrt{\tau}} \int_0^\infty \frac{du}{u^{3/2}}  \frac{\partial}{\partial\tau}\Big\{  \left(\sqrt{\tau + u} - \sqrt{1+\tau + u} -\sqrt{\tau + 2u}+\sqrt{1+\tau + 2u}\ \right) \times \nonumber \\
&\left[ (u+\tau)\ln(u+\tau)-\tau\ln\tau +(1+\tau)\ln(1+\tau)-(1+u+\tau)\ln(1+u+\tau)\right] \Big\} \label{Exrpc} \\
b=&-\frac{ 2}{\pi (-2 + \sqrt{2}) }\int_0^\infty \frac{d\tau}{\sqrt{\tau}} \int_0^\infty \frac{du}{u^{3/2}}  \frac{\partial}{\partial\tau}\Big\{  \left(\sqrt{\tau + u} - \sqrt{1+\tau + u} -\sqrt{\tau + 2u}+\sqrt{1+\tau + 2u}\ \right) \times \nonumber \\
&\left[ (u+\tau)\ln(u+\tau)-(1+u+\tau)\ln(1+u+\tau)-(2u+\tau)\ln(2u+\tau)+(1+2u+\tau)\ln(1+2u+\tau)\right] \Big\}\label{Exrpb}
\end{align}
To comment on the value of $\Omega$, let us remind that 
\begin{align}
\rho(\xi,\xi)=\xi^{\beta}=\xi^{2H-\theta}=\xi^{3H-1- \varepsilon\delta_1+...}=\sqrt{\xi} + \varepsilon(3-\delta_1)\sqrt{\xi}\ln\xi +\mathcal{O}(\varepsilon^2)\label{Arg2}.
\end{align}
If our theory is consistent the coefficient of $\varepsilon\sqrt{\xi}\ln\xi$ in the above equation is set by $\delta_1$, whatever the value of $\delta_2$. The fact that we find the same value with the above argument (\ref{Arg2}) or by the complete calculation (\ref{OmegaComplete}) argues in favor of the consistency of our approach. We also see that there is no term $\varepsilon\sqrt{\xi}$ at first order in the above expression (\ref{Arg2}), which imposes to set $\alpha=0$; this equality can be satisfied by adjusting the value of $\delta_2$. The result is 
\begin{align}
\delta_2=a_1\lambda(a_2+\lambda), 
\end{align}
with the constants 
\begin{align}
a_1 =  -2 (-1 + \sqrt{2}) (4 - 4 \sqrt{2} + b - 6 \ln2  + \sqrt{2} \ln 4) \simeq 1.77, \hspace{1cm} a_2=\frac{-4 + c - \sqrt{2} \ln8}{4 - 4 \sqrt{2} + b - 6 \ln2  + \sqrt{2} \ln 4}\simeq 1.28,
\end{align}
where the numerical values necessitate the numerical evaluation of the integrals in Eqs.~(\ref{Exrpc}) and (\ref{Exrpb}). To summarize, our result at second order for the quenched fBm is 
\begin{align}
&\theta=1-H+\varepsilon \delta_1+\varepsilon^2\delta_2, \hspace{1cm}  \delta_1=- 2 \lambda (\sqrt{2}-1) \hspace{1cm} \delta_2= 1.77 \lambda(1.28 + \lambda) \hspace{1cm}\lambda=1-T.
\end{align}  

\subsection{Second order for the fBm constrained on its past}
For the fBm constrained on its past, we have again to evaluate the second member $C$ in Eq.~(\ref{StartPertTheory}), which reads: 
\begin{align}
C(v,v')&= \int_0^L \frac{du}{2u^{H+\theta}}   \left[\sigma(u+v,u+v')-\frac{\sigma(u+v,u)\sigma(u+v',u)}{\sigma(u,u)} -\sigma(v,v')\right]+ \int_L^\infty \frac{du}{2u^{H+\theta}}   \left[\sigma_s(v,v') -\sigma(v,v')\right].
\end{align}
Now, since $\sigma=\sigma_s$ at first order, and $\theta=1-H+\varepsilon^2\delta_2$, we get:
\begin{align}
C(v,v') &= \int_0^L \frac{du}{2u^{H+\theta}}   \left[\sigma(u+v,u+v')-\frac{\sigma(u+v,u)\sigma(u+v',u)}{\sigma(u,u)} -\sigma(v,v')\right]+ \frac{L^{-\varepsilon^2 \delta_2}}{2\varepsilon^2\delta_2} \left[\sigma_s(v,v') -\sigma(v,v')\right] \nonumber\\
&\simeq -  \frac{\omega_2(v,v')}{2\delta_2}.
\end{align}
This means that the equation for $\rho_0$ is essentially the same as in the first order case, Eq.~(\ref{EqFirstOrder}) as soon as one  replaces $\omega_1/\delta_1$ by $\omega_2/\delta_2$. From Eq.~(\ref{GenSol1}) we deduce   the solution for $\rho_0$ 
\begin{align}
\rho_0(\xi,\xi')=\frac{1}{2\pi\delta_2} \int_0^\infty\frac{dt}{\sqrt{t}}[\omega_2^*(t+\xi,t+ \xi') -\omega_2^*(t+\xi,t ) -\omega_2^*(t ,t+ \xi') +\omega_2^*(t,t)]. \label{9572}
\end{align}
We can identify $\omega_2$ by expanding (\ref{EqSigmaPast}) at second order 
\begin{align}
\omega_2(t,t')
=  - \int_0^\infty dx \frac{(t+x t')\ln(t+xt')-t \ln t - (t'x) \ln (t'x)}{  x (1+x)}.
\end{align}
The integral can be performed after taking the derivatives with respect to $t$ and $t'$ 
\begin{align}
\omega_2^*(t,t')=(\partial_t+\partial_{t'})\omega_2 =-    \frac{1}{6} (2\pi^2  + 3 \ln^2(t'/t)).
\end{align}
Inserting this value into Eq.~(\ref{9572}) leads to
\begin{align}
\rho_0(\xi,\xi')=  \frac{2 }{ \delta_2}   \left[ (\sqrt{\xi} + \sqrt{\xi'}) \ln (\sqrt{\xi}+\sqrt{\xi'}) -\sqrt{\xi} \ln \sqrt{\xi}-  \sqrt{\xi'} \ln \sqrt{\xi'} \right] = \frac{1}{4\pi\delta_2}   \sqrt{  \xi} \ z(\xi/\xi')
\end{align}
 with 
\begin{align}
z(x)=\frac{\pi}{\sqrt{x}}[8 (1 + \sqrt{x}) \ln (1+\sqrt{x}) -4 \sqrt{x} \ln x]\label{zPastFBMPert}
\end{align}
Strikingly we observe a logarithmic divergence of $z(x)$ around $x=0$, as $z(x)\propto 1- 2\ln x$. 
Again, the argument $\rho_0(\xi,\xi)=\sqrt{\xi}$ enables us to identify the value of $\delta_2 = 4 \ln 2$, and thus the result of our perturbation theory for the fBm constrained on its past reads
\begin{align}
\theta\simeq 1-H+ (4\ln 2) \varepsilon^2+...
\end{align}

\section{Result in the limit of high initial temperature, $T=\infty$}
\label{InfiniteT}
Let us consider the quenched fBm in the limit $T\to\infty$, which corresponds physically to the case that the initial temperature is much larger than the temperature of the dynamics. Thus, this corresponds to models starting with a fully ``disordered'' state. We see that the limiting value of the scaling function becomes 
\begin{align}
G_{\infty}(u)\equiv G_{T=\infty}(u)=\frac{-(1+u)^{2H}+ (1+u^{2H}) }{[- 2^{2H}+2 ]u^{2H}}. 
\end{align}
thus we see that the behavior of $G_{T=\infty}(u)$ near $u=1$ is regular. Hence, according to Eq.~(\ref{Behavior_a}), after the Lamperti transform we obtain a smooth Gaussian process whose density of zero crossing can be analyzed with the independent interval approximation.  Denoting again by $a(T)$ the correlator of the process after Lamperti transform, $a_\infty(T)=e^{-HT}G_\infty(e^{-HT})$, 
the result of the IIA is that $\theta$ is the smallest positive number that satisfies the equation $F(-\theta)=0$, where
\begin{align}
F(s)=1+\frac{\pi \ s }{2\vert a_\infty''(0)\vert}\left[1-\frac{2s}{\pi}\int_0^\infty dT e^{-sT} \text{asin}(a_\infty(T))\right]. 
\end{align}
The value of $\theta$ deduced from this approximation for $T=\infty$ are close to $\theta\simeq 0.11\pm 0.01$ for all values of $H$. This means that the perturbation theory around $H=1/2$ will fail at high temperatures, since the weakly non-Markovian limit and the limit of large initial temperature cannot be inverted.




\end{document}